\documentclass[printer]{aa}
\usepackage{graphics,epsfig,lscape,txfonts}
\usepackage{natbib}
\usepackage{german}
\bibpunct{(}{)}{;}{a}{}{,}

\def\cm-2{cm$^{-2}$}

\def\ein{{\it Einstein}}
\def\chandra{{\it Chandra}}

\def\xmm{{XMM-Newton}}
\def\n253{\object{NGC~253}}
\def\m31{\object{M~31}}
\def\me33{\object{M~33}}
\def\mx7{\object{M~33~X$-$7}}
\def\x7{\hbox{X$-$7}}

\newcommand{\ergcm}[1]{$\times 10^{#1}$ \hbox{erg cm$^{-2}$ s$^{-1}$}}

\newcommand{\ergs}[1]{$\times 10^{#1}$ \hbox{erg s$^{-1}$}}
\newcommand{\oergs}[1]{$10^{#1}$ erg s$^{-1}$}
\newcommand{\hcm}[1]{$\times 10^{#1}$ cm$^{-2}$}

\newcommand{\expo}[1]{$\times 10^{#1}$}

\newcommand{\nh}{\hbox{$N_{\rm H}$}}

\begin{document}
\originalTeX

   \title{\xmm\ survey of M~31\thanks{XMM-Newton is an ESA Science Mission 
    with instruments and contributions directly funded by ESA Member
    States and the USA (NASA).}\fnmsep\thanks{Table~2 is available in electronic form
    at the CDS via anonymous ftp to cdsarc.u-strasbg.fr (130.79.128.5)
    or via http://cdsweb.u-strasbg.fr/cgi-bin/qcat?J/A+A/ }
}

   \author{W.~Pietsch \and 
           M.~Freyberg \and
	   F.~Haberl
          }
\institute{Max-Planck-Institut f\"ur extraterrestrische Physik, 85741 Garching, Germany \\
           e-mail: {\tt wnp@mpe.mpg.de}
           }
     

   \date{Received 10 October 2004 / Accepted 6 December 2004}
   \titlerunning{\xmm\ survey of M~31}

	\abstract{ In an analysis of \xmm\ archival observations of the bright 
Local Group spiral galaxy
\m31\ we study the population of X-ray sources (X-ray binaries, supernova
remnants) down to a 0.2--4.5 keV luminosity of 4.4\ergs{34}. EPIC 
hardness ratios and optical and radio information are used to distinguish between 
different source classes. The survey detects in an area of 1.24 square degree 856 sources.
We correlate our sources with earlier \m31\ X-ray catalogues and 
use information from optical, infra-red and 
radio wavelengths.
As \m31\ sources we detect 21 supernova remnants (SNR) and 23 SNR candidates,
18 supersoft source (SSS) candidates, 7 X-ray binaries (XRBs) and 9 XRB candidates, as well as 27 globular 
cluster sources (GlC) and 10 GlC candidates, which most likely are low mass 
XRBs within the GlC. Comparison to earlier X-ray surveys reveal transients not 
detected with \xmm, which add to the number of \m31\ XRBs.
There are 567 sources classified as hard, which may either be 
XRBs or Crab-like SNRs in \m31\ or background AGN. The number of 44 SNRs and 
candidates more than 
doubles the X-ray detected SNRs. 22 sources are new SNR candidates in \m31\
based on X-ray selection criteria.
Another SNR candidate may be the first plerion detected outside the Galaxy and the
Magellanic Clouds.
On the other hand,
six sources are foreground stars and 90 foreground star candidates, 
one is a BL Lac type active galactic nucleus (AGN) and 36 are AGN candidates, 
one source coincides with the Local Group galaxy M 32, 
one with a background galaxy cluster (GCl) and another is a GCl 
candidate, all sources not connected to \m31. 
	
} 
\maketitle

\section{Introduction}

In the \xmm\ survey of the Local Group Sc galaxy M~33,
\citet[][hereafter Paper I]{2004A&A...426...11P}
detected 408 sources in a 0.8 square degree field combining the counts of all
EPIC instruments. The use of X-ray colours and optical and radio information 
allowed them to identify and classify the X-ray sources and proved to be 
efficient in separating super-soft X-ray sources (SSSs) and  thermal supernova 
remnants (SNRs) in  M~33 from Galactic stars in the foreground and ``hard" 
sources. These hard sources may be either X-ray binaries (XRBs) or Crab-like 
SNRs in M~33 or active galactic nuclei (AGN) in the background of the galaxy. 

The Andromeda galaxy \m31\ is located at a similar distance as \me33\ 
\citep[780 kpc,][ i.e.
1\arcsec\ corresponds to 3.8 pc and the flux to  luminosity conversion factor
is 7.3\expo{49} cm$^2$]{1998AJ....115.1916H,1998ApJ...503L.131S} and -- compared
to the near face-on view of \me33\ -- is seen under
a higher inclination (78$^{\circ}$). 
The optical extent of the massive SA(s)b galaxy
can be approximated by an inclination-corrected $D_{25}$
ellipse with large diameter  of 153\farcm3 and axes ratio of 3.09
\citep{1991trcb.book.....D,1988ngc..book.....T}. With its moderate Galactic
foreground absorption  \citep[\nh = 7\hcm{20},
][]{1992ApJS...79...77S} \m31\ is well suited to study the X-ray source
population and diffuse emission in a nearby spiral similar to the Milky Way. 
As \m31\ is seen through a comparable Galactic absorbing column as \me33\
one can use the same methods and similar source selection criteria that 
proved to be successful in the M~33 analysis. On the other hand, there is a big
number of low mass XRBs (LMXBs) identified as bright X-ray sources in \m31\ from
earlier observations (see below). The properties of these sources may help to
better classify some of the ``hard" sources in \me33.

\m31\ was a target for many previous imaging X-ray missions. The \ein\
X-ray observatory detected 108 individual X-ray sources brighter than 
$\sim$5\ergs{36}, 16 of which were 
found to vary between \ein\ observations \citep{1979ApJ...234L..45V,
1990ApJ...356..119C,1991ApJ...382...82T}. 
The sources were identified with young stellar associations, globular 
clusters (i.e. LMXBs) and SNRs \citep[see e.g.][]{1981ApJ...247..879B,
1984ApJ...284..663C}. With the ROSAT HRI, \citet{1993ApJ...410..615P} reported
86 sources brighter than $\sim$\oergs{36} in the central area of \m31, nearly
half of which were found to vary when compared to previous \ein\ observations.
With a separation of about one year between the \m31\ surveys, the ROSAT PSPC covered 
the entire galaxy twice and detected altogether 
560 X-ray sources down to a limit of $\sim$5\ergs{35} 
\citep[][hereafter SHP97, SHL2001]{1997A&A...317..328S,2001A&A...373...63S}. 
The intensity of 34 of the sources varied significantly between the 
observations, and SSSs were established as a new class 
of \m31\ X-ray sources \citep[see also][]{1999A&A...344..459K}.

With the new generation of X-ray observatories, \chandra\ and \xmm, up to now only
parts of \m31\ were surveyed. Deep \chandra\ ACIS-I and HRC observations of the central 
region (covered areas of 0.08 and 0.27 square degree) resolved 204 and 142 X-ray sources, respectively 
\citep{2000ApJ...537L..23G,2002ApJ...577..738K,2002ApJ...578..114K}. 
 Three \m31\ disk fields, spanning a range of stellar populations, were covered
with short \chandra\ ACIS-S observations to compare their point source luminosity
functions to that of the galaxy's bulge \citep{2003ApJ...585..298K}.
A synoptic study of \m31\ with the \chandra\ HRC covered in 17 epochs 
``most" of the disk (0.9 square degree) in short observations and resulted 
in mean 
fluxes and long-term light curves for the 166 objects detected 
\citep{2004ApJ...609..735W}.
In these observations, several \m31\ SNRs were spatially resolved 
\citep{2002ApJ...580L.125K,2003ApJ...590L..21K} and bright XRBs in globular 
clusters and SSSs and quasisoft sources (QSSs) could be characterized 
\citep[][]{2002ApJ...570..618D,2004ApJ...610..247D,2004ApJ...610..261G}.
During the \xmm\ guaranteed time program there were four observations of the central 
area of  \m31\ and three aimed at the northern and two at the southern disk. 
These observations were 
used to investigate the bright and variable sources and diffuse emission
\citep[][]{2001A&A...365L.195S,2001A&A...378..800O,2001ApJ...563L.119T,TKP2004}, 
and to derive source luminosity distributions
\citep{2002ApJ...571L..17T}. In addition, the time variability and spectra of 
several individual XRBs have been studied in detail 
\citep[e.g.][]{2002ApJ...581L..27T,2003A&A...411..553B,2003A&A...405..505B,2004A&A...419.1045M}.

Here we present X-ray images and a source catalogue for the
archival  observations of \m31\ combining the three \xmm\ EPIC instruments and using
only times of low background. For  the source catalogue and source population
study of these observations we  analyzed the individual pointings and the
merged data of the central area simultaneously in five energy bands  in a 
similar way as described in our \me33\ analysis in Paper I. The covered area 
of 1.24 square degree and limiting sensitivity is a significant improvement
compared to the \chandra\ surveys, however, up to now only covers about 2/3 
of the optical \m31\ extent ($D_{25}$ ellipse) with a rather inhomogeneous 
exposure.

\section{Observations and data analysis}
\begin{table*}
\scriptsize
\begin{center}
\caption[]{\xmm\ log of archival \m31\ observation overlapping with the optical 
$D_{25}$ ellipse (proposal numbers 010927, 011257 and 015158).}
\begin{tabular}{lllrrrlrlrlr}
\hline\noalign{\smallskip}
\hline\noalign{\smallskip}
\multicolumn{1}{c}{M 31 field} & \multicolumn{1}{c}{Obs. id.} &\multicolumn{1}{c}{Obs. dates} &
\multicolumn{2}{c}{Pointing direction} & \multicolumn{1}{c}{Offset~$^*$} & \multicolumn{2}{c}{EPIC PN} & 
\multicolumn{2}{c}{EPIC MOS1} & \multicolumn{2}{c}{EPIC MOS2}  \\ 
\noalign{\smallskip}
& & & \multicolumn{2}{c}{RA/DEC (J2000)} & \multicolumn{1}{c}{} 
& \multicolumn{1}{c}{Filter$^{+}$}  & \multicolumn{1}{c}{$T_{exp}^{\dagger}$}
& \multicolumn{1}{c}{Filter$^{+}$}  & \multicolumn{1}{c}{$T_{exp}^{\dagger}$}
& \multicolumn{1}{c}{Filter$^{+}$}  & \multicolumn{1}{c}{$T_{exp}^{\dagger}$}\\
\noalign{\smallskip}
\multicolumn{1}{c}{(1)} & \multicolumn{1}{c}{(2)} & \multicolumn{1}{c}{(3)} & 
\multicolumn{1}{c}{(4)} & \multicolumn{1}{c}{(5)} & \multicolumn{1}{c}{(6)} & 
\multicolumn{1}{c}{(7)} & \multicolumn{1}{c}{(8)} & \multicolumn{1}{c}{(9)} & 
\multicolumn{1}{c}{(10)} & \multicolumn{1}{c}{(11)} & \multicolumn{1}{c}{(12)}\\
\noalign{\smallskip}\hline\noalign{\smallskip}
Centre 1~(c1) & 0112570401 & 2000-06-25    & 0:42:36.2 & 41:16:58 & $-1.9,+0.1$ & medium  & 26.40(23.60) & medium  & 29.92(29.64) &medium  & 29.91(29.64) \\
Centre 2~(c2) & 0112570601 & 2000-12-28    & 0:42:49.8 & 41:14:37 & $-2.1,+0.2$ & medium  &  9.81(~~5.85) & medium  & 12.24(~~6.42) &medium  & 12.24(~~6.42) \\
Centre 3~(c3) & 0112570701 & 2001-06-29    & 0:42:36.3 & 41:16:54 & $-3.2,-1.7$ & medium  & 27.65(21.83) & medium  & 27.65(23.85) &medium  & 27.65(23.86) \\
North 1~~\,(n1) & 0109270701 & 2002-01-05    & 0:44:08.2 & 41:34:56 & $-0.3,+0.7$ & medium  & 54.78(48.57) & medium  & 57.31(55.68) &medium  & 57.30(55.67) \\
Centre 4~(c4) & 0112570101 & 2002-01-06/07 & 0:42:50.4 & 41:14:46 & $-1.0,-0.8$ & thin	& 60.79(48.11) & thin	 & 63.31(52.87) &thin	 & 63.32(52.86) \\
South 1~~\,(s1) & 0112570201 & 2002-01-12/13 & 0:41:32.7 & 40:54:38 & $-2.1,-1.7$ & thin	& 53.45(47.00) & thin	 & 53.76(51.83) &thin	 & 53.73(51.84) \\
South 2~~\,(s2) & 0112570301 & 2002-01-24/25 & 0:40:05.0 & 40:34:38 & $-1.1,-0.3$ & thin	& 29.31(22.23) & thin	 & 29.37(24.23) &thin	 & 29.38(24.24) \\
North 2~~\,(n2) & 0109270301 & 2002-01-26/27 & 0:45:28.6 & 41:55:26 & $-0.3,-1.5$ & medium  & 26.27(22.86) & medium  & 26.54(25.22) &medium  & 26.55(25.28) \\
North 3~~\,(n3) & 0109270401 & 2002-06-29/30 & 0:46:31.1 & 42:17:32 & $-2.3,-1.7$ & medium  & 51.39(41.28) & medium  & 59.69(45.39) &medium  & 59.69(45.53) \\
Halo 4~~~~(h4) & 0151580401 & 2002-02-06    & 0:46:15.9 & 41:20:25 & $+0.3,+0.0$ & medium  & 11.29(11.29) & medium  & 12.91(12.91) &medium  & 12.92(12.92) \\
\noalign{\smallskip}
\hline
\noalign{\smallskip}
\end{tabular}
\label{observations}
\end{center}
Notes:\\
$^*~~~$: Systematic offset in Ra and DEC in arcsec determined from correlations with 2MASS, USNO-B1 and \chandra\ catalogues \\
$^{ +~}$: all observations in full frame imaging mode\\
$^{ {\dagger}~}$: Exposure time in units of ks after screening for high
background used for detection, for colour image in brackets (see text)\\
\normalsize
\end{table*}

Table~\ref{observations} summarizes the archival \xmm\
\citep{2001A&A...365L...1J}  EPIC 
\citep{2001A&A...365L..18S,2001A&A...365L..27T} observations which at least
partly overlap with the inclination corrected optical \m31\ $D_{25}$ ellipse.
For each observation we give the field name and the abbreviation used in the
text for the \m31\ observation  (Col. 1),  the observation identification (2),
date (3), pointing direction (4,5) and systematic offset (6), 
as well as filter and exposure time after
screening for high background for EPIC PN (7,8), MOS1 (9,10), and MOS2 (11,12).
For creating the  full field colour image we had to apply an even more stringent
background  screening which resulted in the shorter exposure times given in
brackets.

In the \xmm\ observations the EPIC PN and MOS instruments  were operated in
the  full frame mode providing a time resolution of 73.4 ms and 2.6 s,
respectively.  The medium filter was in front of the EPIC cameras in all but
the  observations c4, s1 and s2 which were performed with the thin filter. 
For creating the merged images and source detection we used medium and thin
filter observations together. This procedure can be justified as the difference
in absorption between the medium and soft filters corresponds to an 
\nh = 1.1\hcm{20}, significantly lower than  the Galactic 
foreground absorption.

We carefully screened the event files for bad CCD
pixels remaining after the standard processing.  To create a homogeneous
combined colour image with similar background level for all fields we had to carefully
screen the data for times of high background using the high energy (7--15 keV) background light
curves provided by the SAS  tasks {\tt epchain} and {\tt emchain}. For the images that were used
for the source detection procedures in the individual fields we adopted a less 
stringent background
screening which due to the then longer exposures allowed us to detect fainter
sources. The good
time intervals (GTI) were determined from the higher statistic PN light curves
and also used for the MOS cameras. Outside the PN time coverage, GTIs were
determined from the combined MOS light curves.  The corresponding low
background times for the individual observations are listed for the PN and MOS
cameras in Table~\ref{observations}. 

The archive contains four observations of the centre area of \m31\ separated by
half a year adding up to low background exposures for source detection of
124.6, 133.1 and 133.2 ks for EPIC PN, MOS1 and MOS2 cameras, respectively. In
addition there are two pointings in the southern and three in the northern disk of \m31\
that aimed for 60 ks exposure. Due to high background, for two of the 
observations the time usable for detection is significantly shorter (s2, n2).
In addition the archive contains four shorter pointings into the halo of \m31\
(h1--h4, PI Di Stefano). These
observations mainly contain foreground and background objects as they were
pointed far of the disk of the galaxy. Only observation h4 (11 ks low
background) partly covers the disk and therefore is included in our analysis. 
In total, the observations in our analysis cover an area of 1.24 square degree.

The data are treated in a similar way as the M~33 data described in Paper I.
We used five energy bands: (0.2--0.5) keV, (0.5--1.0) keV, (1.0--2.0) keV,
(2.0--4.5) keV, and (4.5--12) keV as band 1 to 5. We intentionally split the
(0.5--2.0) keV band used in the 1XMM XMM-Newton Serendipitous Source 
Catalogue\footnote{Prepared by the XMM-Newton Survey Science Centre Consortium 
({\tt http://xmmssc-www.star.le.ac.uk/})} to get on average a more homogeneous
distribution of the source counts to the energy bands which leads to a better
spread of the hardness ratio values and allows effective source classification
(see Paper I and Sect. 5).

For PN we selected only ``singles" (PATTERN = 0) in band 1, for the other bands
``singles and doubles" (PATTERN $\le$ 4). For MOS we used ``singles" to
``quadruples"  (PATTERN $\le$ 12). To avoid background variability over the PN
images we omitted the energy range (7.2--9.2) keV in band 5 where strong
fluorescence lines cause higher background in the outer detector area
\citep{2004SPIE.5165..112F}. To convert source count rates in the individual
bands to fluxes we used count rate to energy conversion factors (ECF) for
PN and MOS observations calculated for the epoch of observation and filters
used, assuming the same spectrum as for the first \xmm\ source catalogue, i.e. 
a power law spectrum with photon index 1.7 absorbed by the Galactic foreground 
column of 7\hcm{20}.  As has been demonstrated in Table 2 of Paper I, ECF
values only vary by about 20\% for PN in energy bands 1 to 4 
for different spectra like an absorbed 1 keV thin
thermal model typical for SNRs or a 30~eV black body model typical for SSSs. The
same is true for EPIC MOS, with the exception of the black body ECF in the
0.2--0.5 keV band where the MOS sensitivity is much lower.  For the detection in the
merged centre images (combining observations with thin and medium filter)  
we used averaged ECFs which may lead to wrong flux estimates of at maximum 5\%. 

For most sources band 5 just adds noise to the total count rate. If converted
to  fluxes due to the high ECF this noise often dominates the total flux. To
avoid this  problem we calculated count rates and fluxes for detected sources
in the ``XID" (0.2--4.5) keV band (bands 1 to 4 combined).  While for most sources this is
a good solution for extremely hard or  soft sources there may still be bands
just adding noise. This then leads to rate  and flux errors that seem to
wrongly indicate a lower source significance. A  similar effect occurs for the
all instrument rates and fluxes if a source is  mainly detected in one
instrument (e.g. soft sources in PN).

To classify the source spectra we computed four hardness ratios from the source
count rates. These hardness ratios and errors are defined as  

$HRi = \frac{B_{i+1} - B_{i}}{B_{i+1} + B_{i}}$  and
$EHRi = 2  \frac{\sqrt{(B_{i+1} EB_{i})^2 + (B_{i} EB_{i+1})^2}}{(B_{i+1} + B_{i})^2}$, 

for {\it i} = 1 to 4, 
where $B_{i}$ and $EB_{i}$ denotes count rates and corresponding errors in
band {\it i} as defined above.  In the standard source detection products  hardness
ratios and count rates in individual energy bands are not combined for all
instruments. To improve statistics we calculated for all energy bands ``all
EPIC"
count rates, fluxes and hardness ratios. We are aware that these  products have
some jitter in their meaning, as the relative integration times for  individual
sources in the EPIC instruments differ. Due to the better  signal in
the combined products they are still very valuable for source  classification.
For investigations needing accurate calibration the products for  the
individual instruments should be used that are also given.

We created for PN, MOS1 and MOS2 in each of the 5 energy bands  mentioned above
images, background images, exposure maps (without and with vignetting
correction),  masked them for acceptable detector area. For PN the background
maps contain the contribution from the ``out of time (OOT)" events (parameter
{\tt withootset=true} in task {\tt esplinemap}). In contrast to our \me33\ 
raster, the \m31\ images (with the exception of the four observations towards the
\m31\ centre, c1, c2, c3, c4) only overlap at the edge of the field of view (FOV).  
We therefore searched for sources in each of these fields individually and only 
merged the centre fields for source detection. For visualization purposes we
created images of all observations individually, for the centre and for the full
area merging all EPIC instruments (see Sect. 4).
To allow an easy merging of the galaxy centre images, we re-calculated for 
the events of all contributing observations projected sky coordinates 
with respect to reference position 
RA=00$^{\rm h}$42$^{\rm m}$45\fs0, DEC=$+$41\degr15\arcmin00\arcsec (J2000),
for the full field images with respect to reference position
RA=00$^{\rm h}$43$^{\rm m}$24\fs0, DEC=$+$41\degr24\arcmin00\arcsec (J2000),
respectively.

To create event lists and images we used the latest calibration products 
for the linearisation of the EPIC MOS
CCDs (MOS*\_LINCOORD\_0017.CCF) and EPIC boresight (XMM\_BORESIGHT\_0018.CCF). We verified
with the source-rich \m31\ centre pointings that PN and MOS source positions
coincide to better than 0.5\arcsec. On the other hand there were still significant offsets 
between the observations that had to be corrected for before
merging. For the centre observations these offsets were
determined from source lists of the individual observations. We used the
USNO-B1, 2MASS, and \chandra\ catalogues to define an absolute reference frame.
The offets applied are listed in column 6 of Table~\ref{observations}.
This finally resulted in a residual systematic position error of less than 0.5\arcsec.

The images for PN, MOS1 and MOS2 do not fully
overlap at the borders due to the different FOV of the EPIC 
instruments. Nevertheless,  we used the full area for source detection.

The data analysis was performed using tools in the SAS v6.0 and some later
versions from the development area as specially mentioned,  EXSAS/MIDAS
03OCT\_EXP, and  FTOOLS v5.2 software packages, the imaging application DS9 v3.0b6
and spectral  analysis software XSPEC v11.2.   

\section{Source catalogue}
We searched for sources using simultaneously $5\times 3$
images (5 energy bands and PN, MOS1 and MOS2 camera). For the pointings into 
the disk and halo of \m31\ this procedure was applied to the individual observations. 
The centre pointings strongly overlap and we therefore merged the images to reach 
higher detection sensitivity. A preliminary source
list created with the task {\tt boxdetect} with a low likelihood threshold  was
used as starting point for the task {\tt emldetect} v. 4.40.  To resolve sources that overlap due to the PSF,
we used parameter {\tt multisourcefit=4}.  We selected sources equivalent to a
single camera likelihood of 10 by accepting sources in the combined fit which
have a likelihood above 7. This procedure produced acceptable results when
checked against the smoothed images (see  Sect.~4). Two emission peaks were
resolved into several sources with no sign of  multiplicity from the smoothed
images. In a specific detection run we found these sources (747 and 832) 
with a high likelihood for extent and that they can be characterized by a
King beta model with a core radius of 30\farcs1$\pm$0\farcs3 and 
28\farcs3$\pm$1\farcs2, respectively. 
We rejected spurious detections in extended emission around the centre.
Finally we merged the source lists of the centre and individual pointings
to the disk and ordered the sources according to increasing right ascension. 
For sources detected in more than one observation we included the 
parameters of that source in the catalogue that was found with the 
higher detection likelihood. A recurrent 
transient source was reported from \chandra\ and \xmm\ observations 
\citep{2000ApJ...537L..23G,2001A&A...378..800O,2004ApJ...609..735W}. It is only
visible during the short observation c2 and was not significantly detected above 
the central diffuse emission by the detection procedure in the merged centre 
observation. 
We therefore derived the source parameters of this source
from just observation c2, and added them at the end of the 
catalogue.
In total we detect 856 sources in the field.

The source parameters are summarized in Table~2  
(EPIC combined products
and products for EPIC PN, MOS1 and MOS2, separately) which is available in 
electronic form at the CDS. 

Table~\ref{master} 
gives the source number (Col. 1), detection field, from which the source was entered 
into the catalogue (2),
source position (3 to 9) with $1\sigma$
uncertainty radius (10), likelihood of existence (11), integrated PN,
MOS1 and MOS2 count rate and error (12,13) and flux and error (14,15) in the
(0.2--4.5) keV XID band, 
and hardness ratios and errors (16--23). 
Hardness ratios are calculated only for sources for which at 
least one of the two band count rates has a significance greater than $2\sigma$.  
Errors are the properly combined statistical errors in each band and can
extend beyond the range of allowed values of hardness ratios as defined previously 
(--1.0 to 1.0). 
The EPIC instruments contributing to the source 
detection, are indicated in the ``Val" parameter (Col. 24, first character for 
PN, second MOS1, third MOS2) as ``T", if inside the field of view (FOV), or ``F", 
if outside of FOV. 
There are 52 sources at the periphery of the FOV where only part of the EPIC 
instruments contribute. The positional error (10) does not include intrinsic systematic errors 
which amount to 0\farcs5 (see above) and should be 
quadratically added to the statistical errors. 

Table~2 then gives for EPIC PN exposure (25), source existence 
likelihood (26),  count rate and error (27,28) and flux and error (29,30) in the
(0.2--4.5) keV XID band,
and hardness ratios and error (31--38). Columns 39 to 52 and 53 to 66 give the same information 
corresponding to Cols. 25 to 38, but now for the EPIC MOS1 and MOS2 instruments. Hardness ratios 
for the individual instruments were again screened as described above. From the comparison of the 
hardness ratios derived from integrated PN, MOS1 and MOS2 count rates
(Cols. 16--23) and the hardness ratios of the individual instruments (Cols. 31--38, 45--52 and 59--66) 
it is clear that combining the instrument 
count rate information yielded significantly more hardness ratios above the chosen 
significance threshold.

Column 67 shows cross correlations with \m31\ X-ray catalogues in the literature (see 
detailed discussion in Sect.~5).
Only 364 sources in our catalogue correlate with previously reported \m31\ X-ray 
sources, i.e. our catalogue doubles the number of known X-ray sources in \m31. 

In the remaining columns of Table~2, we give cross correlation information
with sources in other wavelength ranges,
which is further described in Sect.~6. We only want to mention here, that we used the 
foreground stars, globular cluster sources and candidates, to verify the assumed  source 
position errors. All but six of the 33 identifications
and 100 candidates are located within the $3\sigma$ statistical plus systematic 
positional error given above.

The faintest sources detected have an XID band flux of 6.0\ergcm{-16}. The
brightest source (50, XID band flux of
 4.5\ergcm{-12}) is identified as AGN in the \m31\ field. 
The brightest \m31\ source (291, an XRB) has an XID band flux of  3.9\ergcm{-12}. 
The \xmm\ detected sources in \m31\ therefore cover an absorbed luminosity range 
of 4.4\ergs{34} to 2.8\ergs{38} in the XID band.

\section{Images}
\begin{figure*}
   \resizebox{\hsize}{!}{\includegraphics[bb=117 208 496 585,clip]{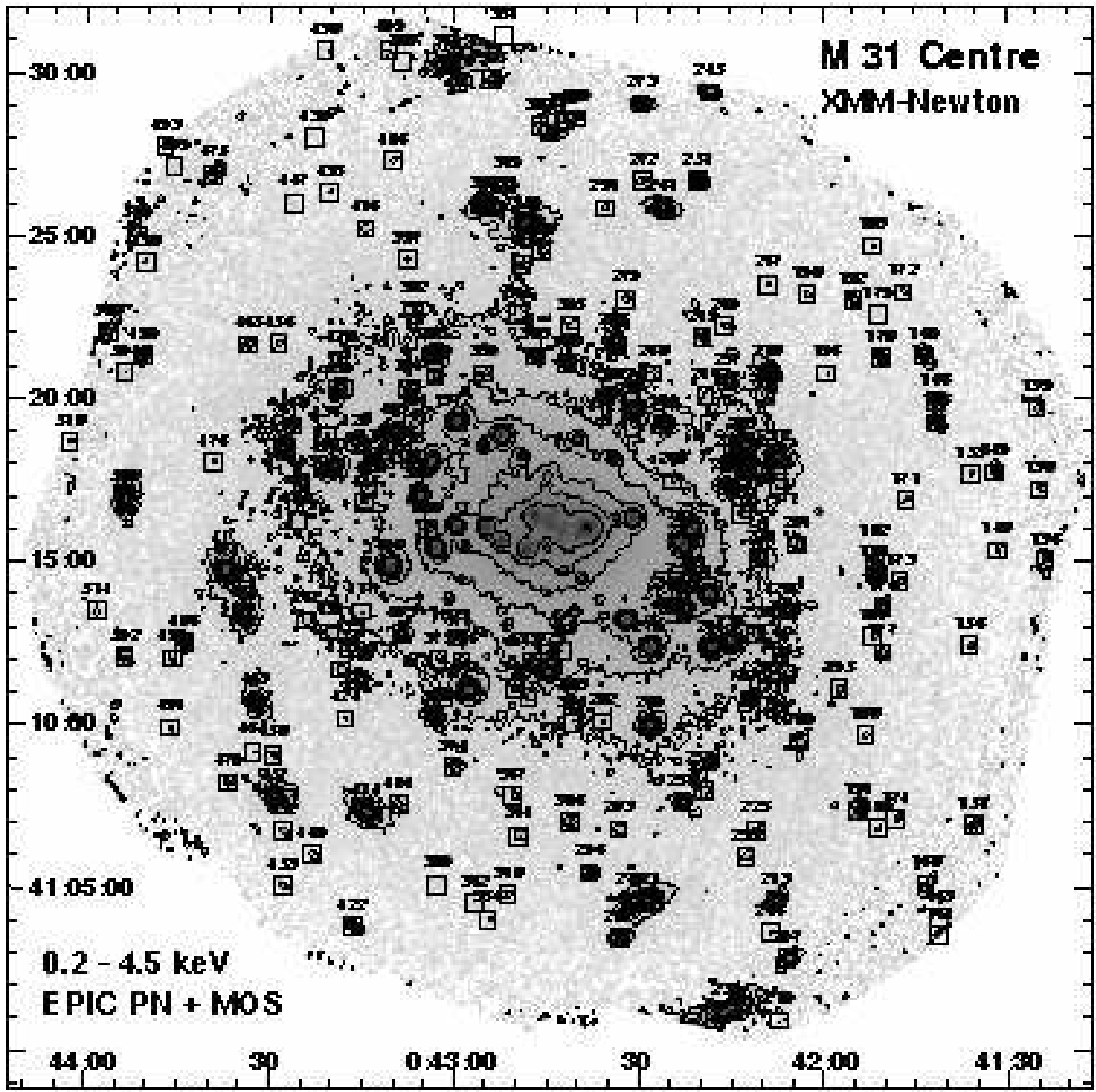}}
    \caption[]{
Logarithmically-scaled \xmm\ EPIC low background image  with a pixel size of 1 arcsec$^2$
of the \m31\ centre observations combining PN and MOS1 and MOS2 
cameras in the (0.2--4.5)~keV XID band. 
The data are smoothed with a Gaussian
of $FWHM$ 5\arcsec\ which corresponds to the point spread function in the centre area of 
the combined observations. The image is 
corrected for un-vignetted exposure and masked for 
     exposures above 5 ks for the individual cameras. 
Contours are at $(1, 2, 4, 8, 16, 32)\times 10^{-6}$ ct s$^{-1}$ pix$^{-1}$ 
including a factor of two smoothing. 
Sources from the catalogue are marked in the outer area. The inner area is shown in detail 
in Fig.~\ref{zoom}.
}
    \label{icentre} 
\end{figure*}

\begin{figure*}
   \resizebox{\hsize}{!}{\includegraphics[bb=120 230 492 562,clip]{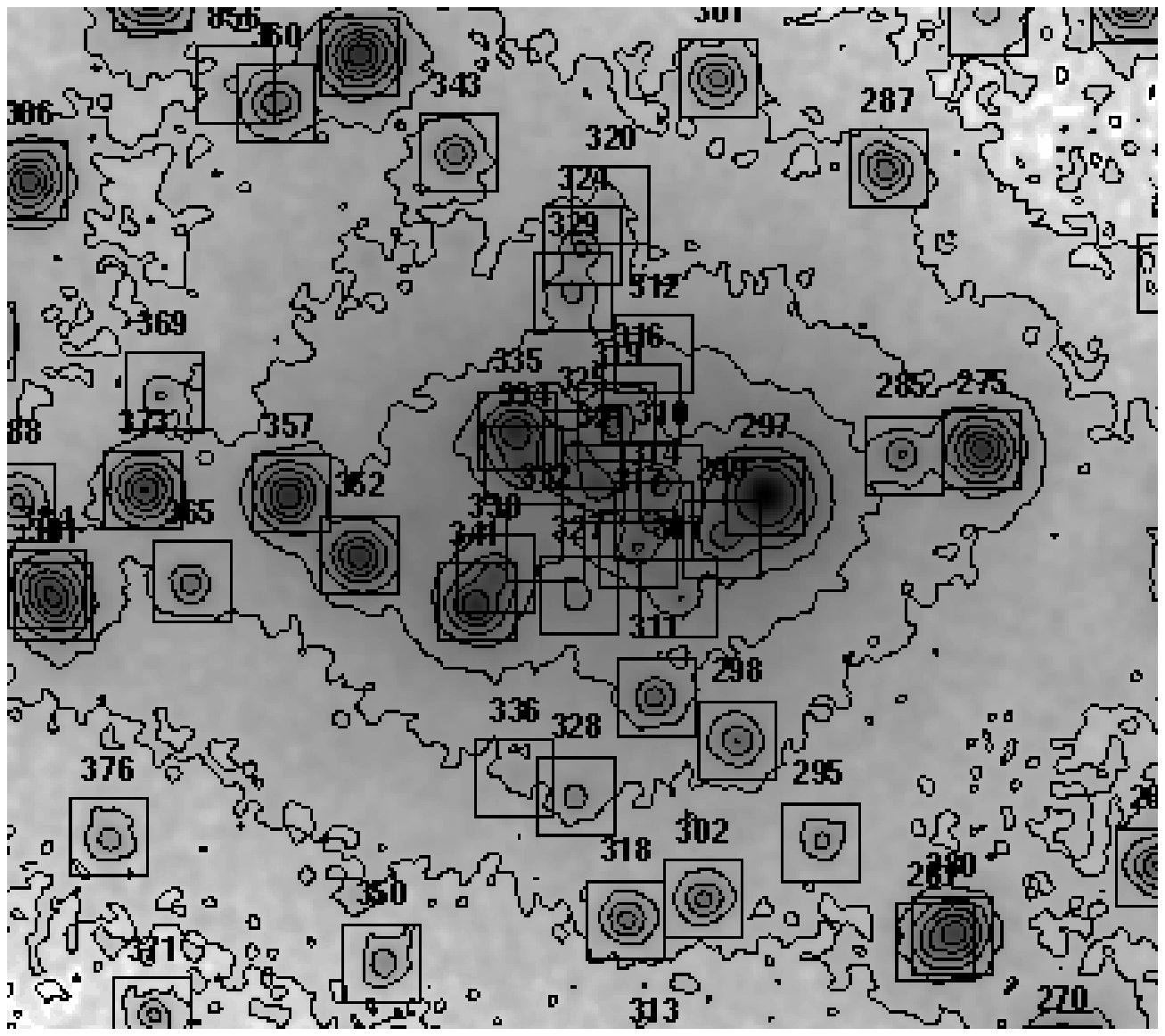}\includegraphics[bb=120 232 490 562,clip]{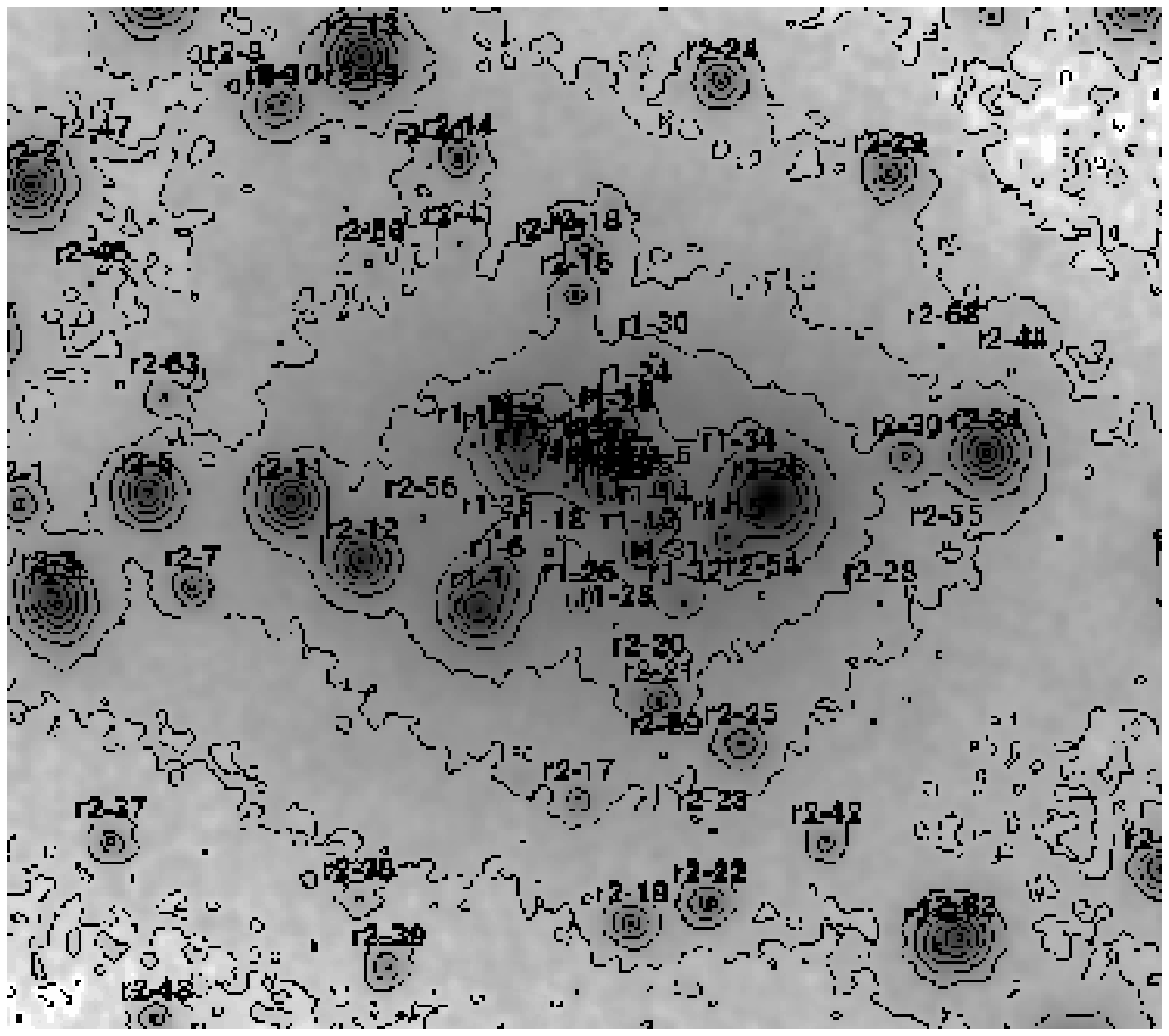}}
    \caption[]{
Inner area of Fig.~\ref{icentre}.
Contours are at $(1, 2, 4, 8, 16, 32, 64, 128, 256)\times 10^{-6}$ ct s$^{-1}$ pix$^{-1}$. 
{\bf Left:} Sources from the \xmm\ catalogue are marked as 30\arcsec$\times$30\arcsec squares. 
{\bf Right:} Sources from the \chandra\ 
analysis of \citet{2002ApJ...577..738K} and \citet{2004ApJ...609..735W} are marked.
}
    \label{zoom} 
\end{figure*}

\begin{figure*}
    \resizebox{\hsize}{!}{\includegraphics[bb=117 208 496 585,clip]{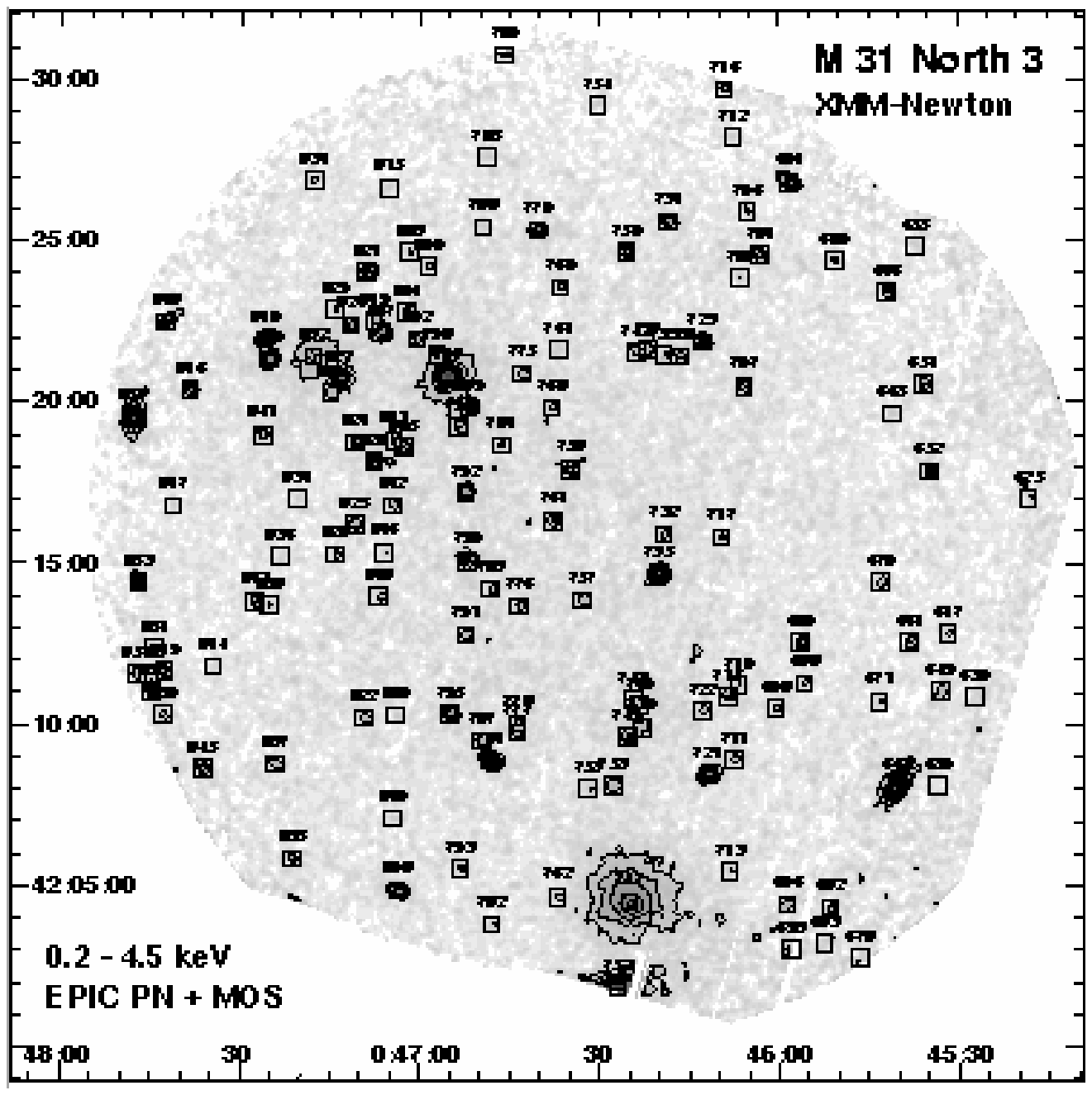}\hskip0.2cm\includegraphics[bb=117 208 496 585,clip]{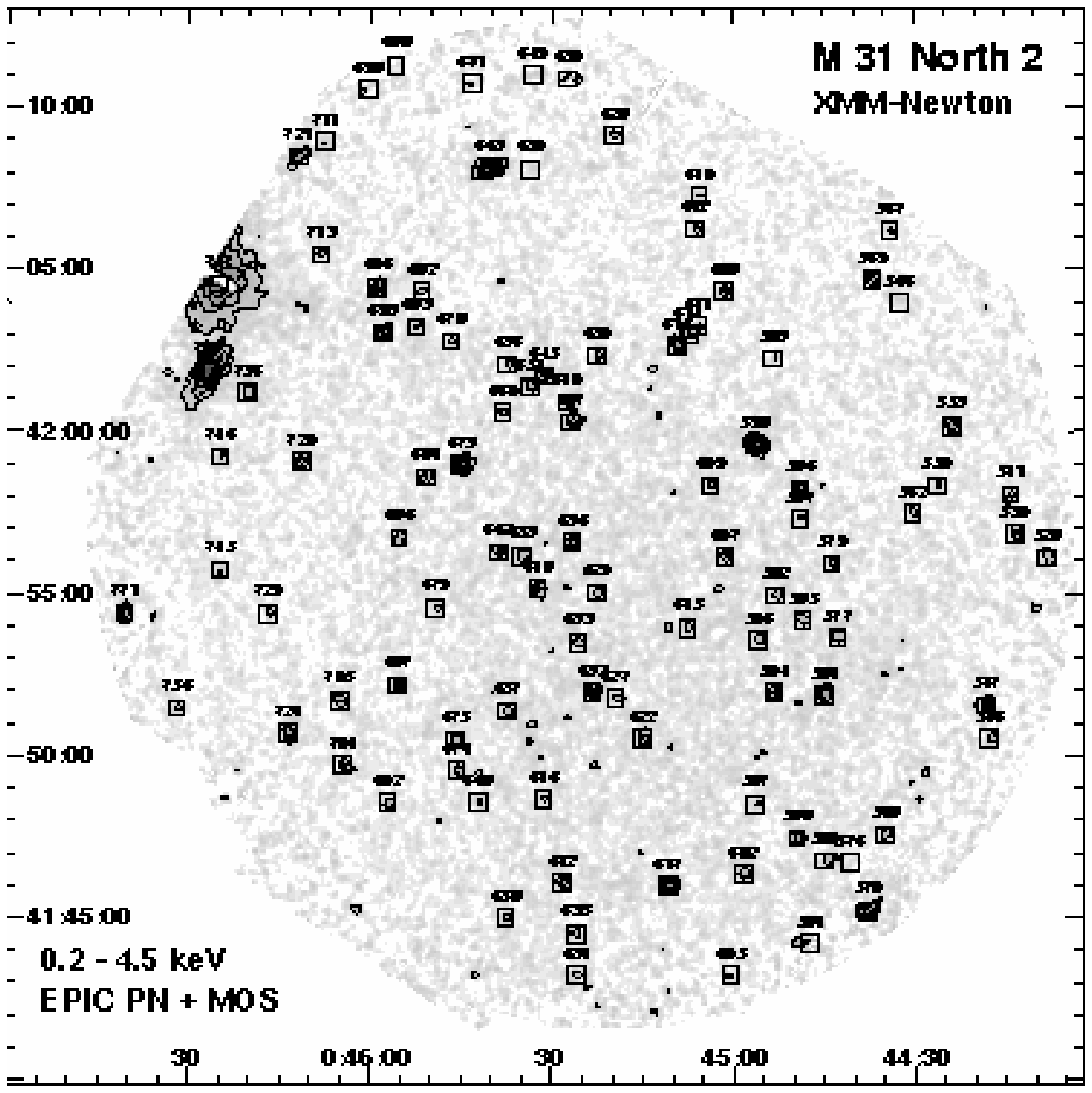}}
        \resizebox{\hsize}{!}{\includegraphics[bb=117 208 496 585,clip]{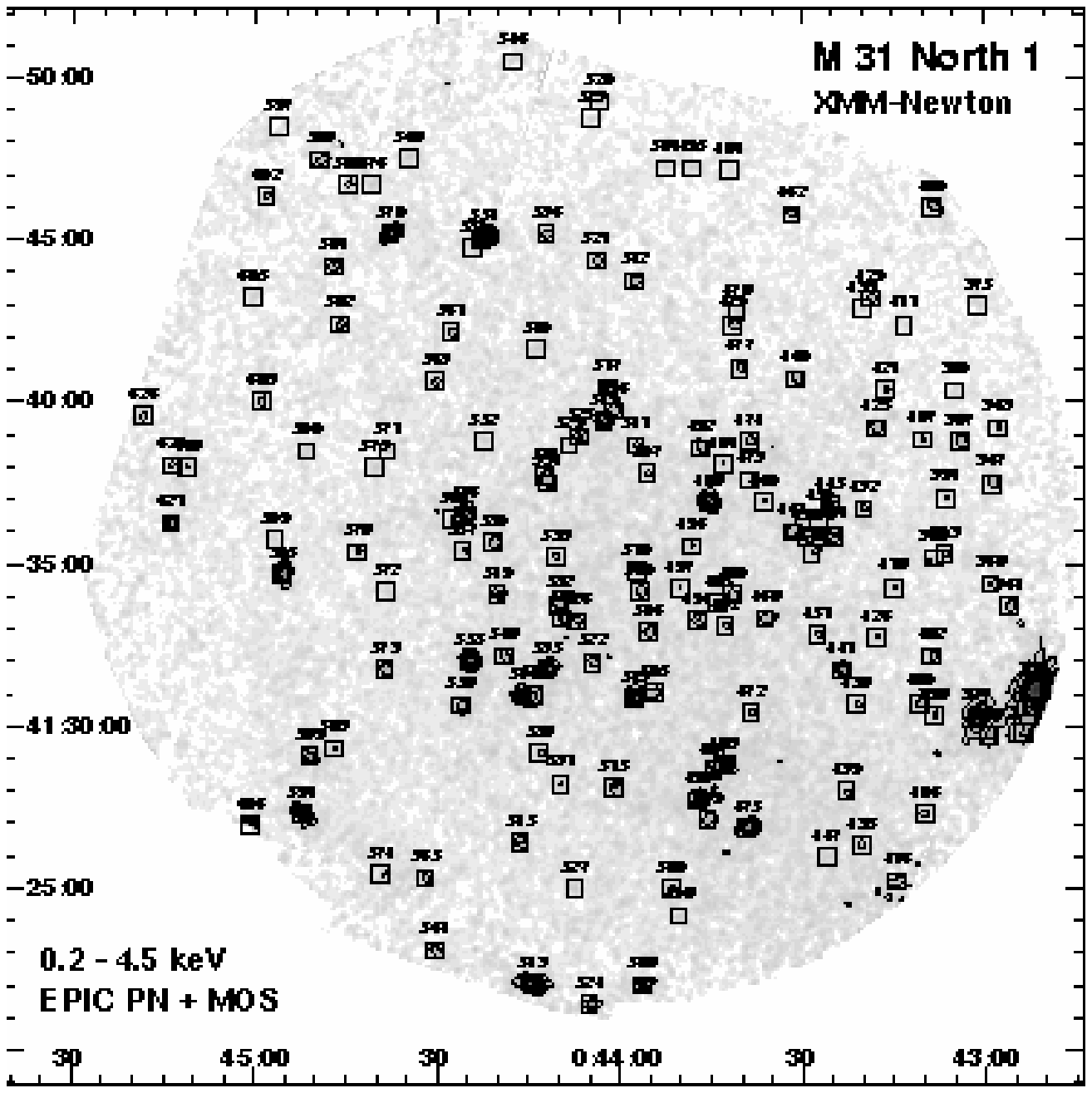}\hskip0.2cm\includegraphics[bb=117 208 496 585,clip]{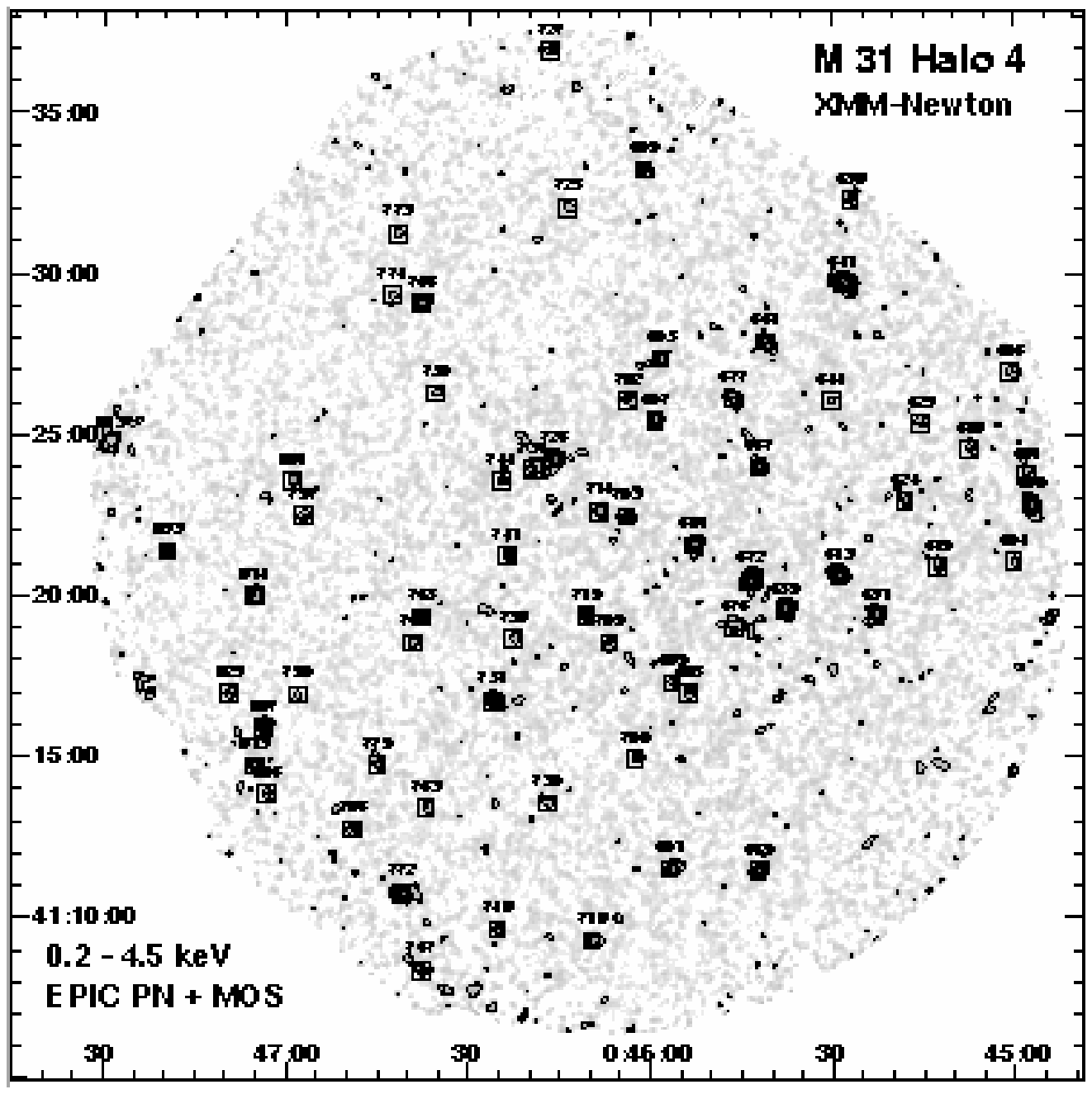}}
	     \caption[]{
     \xmm\ EPIC \m31\ images in the (0.2--4.5) keV XID band: North 3 ({\bf upper left}),
     North 2 ({\bf upper right}),
     North 1 ({\bf lower left}), and 
     Halo 4 ({\bf lower right}).
}
    \label{imax_rest} 
\end{figure*}
\begin{figure*}
    \resizebox{\hsize}{!}{\includegraphics[bb=117 208 496 585,clip]{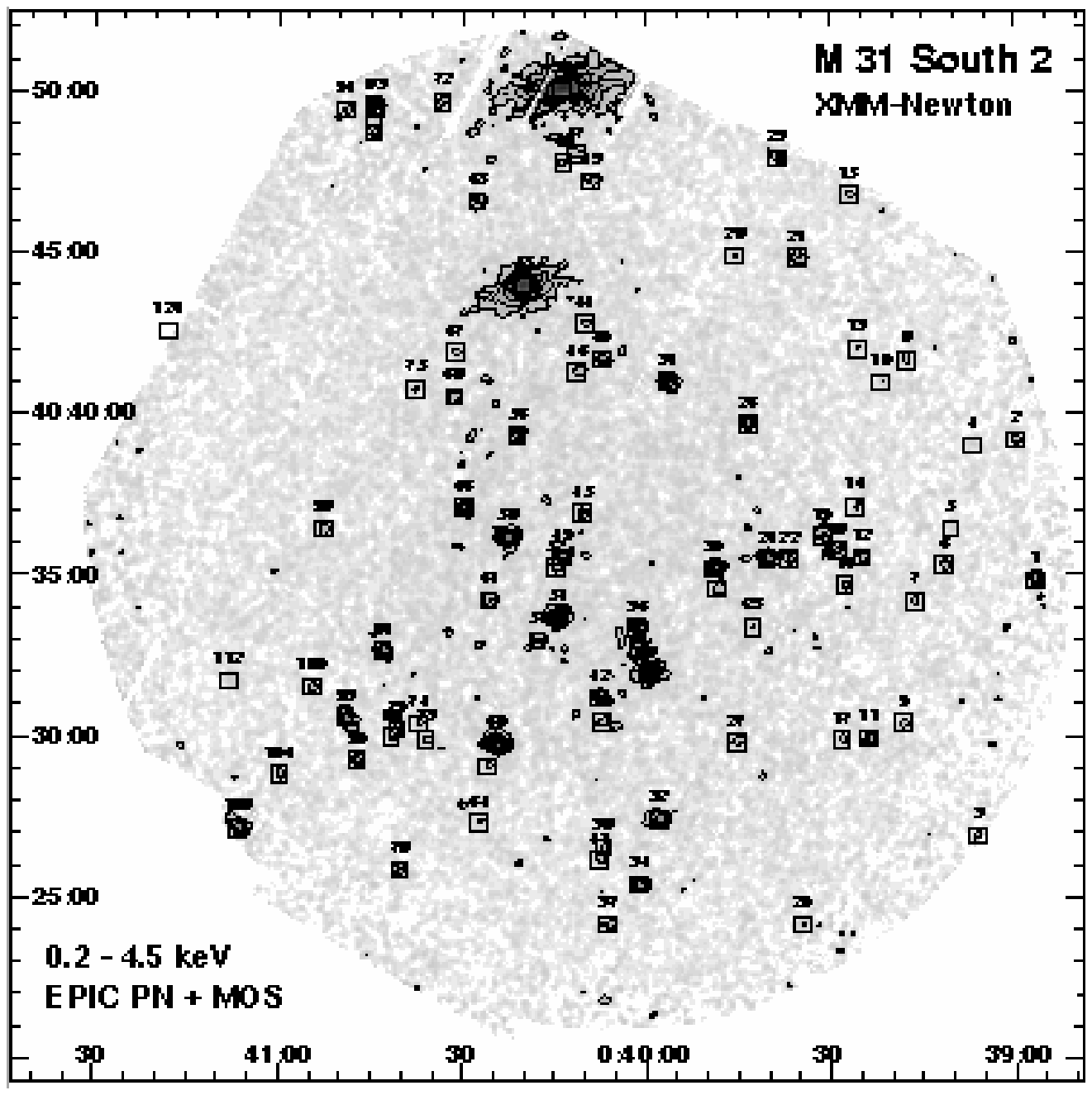}\hskip0.2cm\includegraphics[bb=117 208 496 585,clip]{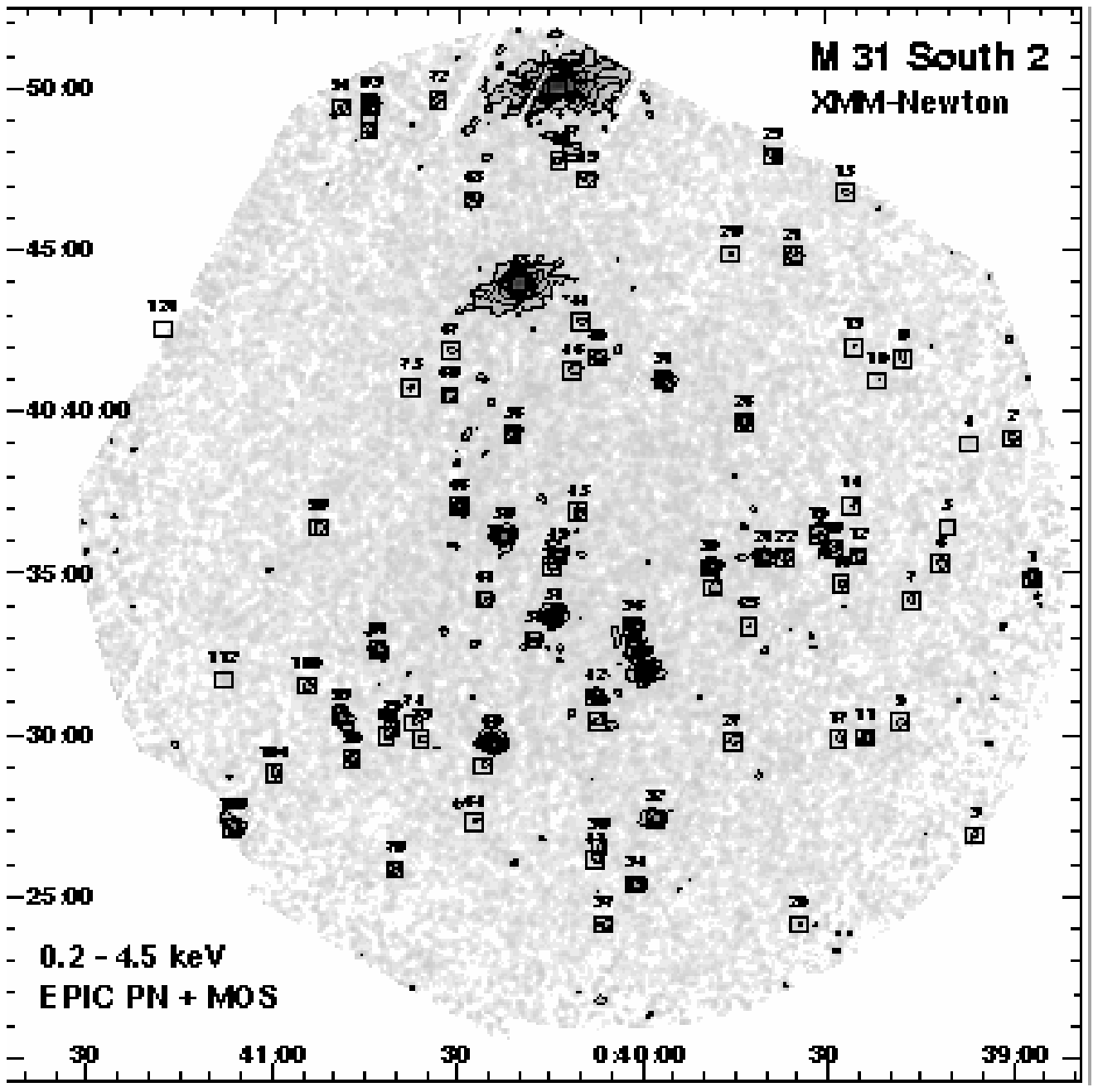}}
    \addtocounter{figure}{-1}
     \caption[]{(continued)
     \xmm\ EPIC \m31\ images in the (0.2--4.5) keV XID band:  
     South 1 ({\bf left}) and
     South 2 ({\bf right}). The images are integrated in 2\arcsec pixels and 
     smoothed with a Gaussian
     of $FWHM$ 10\arcsec, corrected for un-vignetted exposure and masked for 
     exposures above 5 ks for the individual cameras. Contour levels are 
     $(4, 8, 16, 32, 64, 128)\times 10^{-6}$ ct s$^{-1}$ pix$^{-1}$ including a 
     factor of two smoothing. 
     Sources from the catalogue are indicated. 
}
\end{figure*}
\begin{figure*}
  \resizebox{12cm}{!}{
  }
  \hfill
  \parbox[b]{55mm}{
    \caption[]{
Logarithmically-scaled, three-colour \xmm\ EPIC low background image of the \m31\ 
medium and thin filter observations combining PN and MOS1 and MOS2 
cameras. Red, green and
blue show respectively the (0.2--1.0)~keV, (1.0-2.0)~keV 
and (2.0--12.0)~keV bands. 
The data in each energy band have been smoothed with a Gaussian
of $FWHM$ 20\arcsec\ to use an average point spread function for the 
different off-axis angles. The camera images have been 
corrected for un-vignetted exposure and masked for exposure above 5 ks. The 
image scale and the optical $D_{25}$ ellipse of \m31\ are marked. 
Shown in detail to the upper right is a factor three zoom-in to the
bulge area using images binned to 1\arcsec\ and smoothed with a
Gaussian $FWHM$ 5\arcsec\ corresponding to the point-spread-function (PSF)
in the center of the field-of-view. 
}
    \label{rgb}} 
\end{figure*}
For Figs.~\ref{icentre},~\ref{imax_rest},~\ref{rgb}, we smoothed the images and
exposure maps used for detection  with a Gaussian ($FWHM$ of 5\arcsec, 10\arcsec\
and 20\arcsec, respectively). For PN
we subtracted OOT images, that were masked and smoothed in the same way as the
images. For the EPIC combined images we added the images of the individual
cameras scaled according to the background in the individual energy bands. For
the colour image Fig.~\ref{rgb}, we added images for the individual
bands as needed, masked them to a total un-vignetted exposure of 5 ks and 
indicated the optical extent of \m31\ by the $D_{25}$ ellipse. 

The (0.2--4.5) keV XID band images (Figs.~\ref{icentre},~\ref{imax_rest}) 
give an overview on the sources detected in the \xmm\ analysis. To
better visualize faint structures we added contours. 
By comparing images in the different energy bands it is clear that many sources 
only show up
in some of the images, a fact that indicates spectral diversity and is further 
quantified in the different hardness ratios of the sources. This fact can be
visualized even more clearly in the combined EPIC colour image where we coded
the (0.2--1.0) keV band in red, the (1.0--2.0) keV band in green and the (2.0--12) keV band in
blue (see Fig.~\ref{rgb}). The image is a demonstration of the colourful X-ray
sky.  SSSs, thermal SNRs and foreground stars appear red or yellow, XRBs,
Crab-like SNRs and AGN green to blue. Bright diffuse emission fills the bulge 
and fainter emission the area of the disk (yellow and red, see also 
Fig.~\ref{icentre}). A detailed analysis of this emission is outside the scope 
of this paper. \xmm\ analysis of the diffuse emission of the \m31\ bulge region  
was reported by \citet[][]{2001A&A...365L.195S} and \citet{2004ApJ...615..242T}
and of the northern disk by \citet{TKP2004}.

\section{Cross-correlation with other \m31\ X-ray catalogues}
\addtocounter{table}{1}
\begin{table*}
\begin{center}
\caption[]{Summary of cross-correlation with the \m31\ catalogue of 
the first (July 1991) and second (August 1992) ROSAT PSPC survey 
(SSHP97,SHL2001). ROSAT sources, not detected in our \xmm\ analysis, 
are arranged according to detection likelihood (LH) levels.}
\begin{tabular}{lll}
\hline\noalign{\smallskip}
\hline\noalign{\smallskip}
\multicolumn{1}{c}{} & 
\multicolumn{1}{c}{SSHP97 source numbers} &
\multicolumn{1}{c}{SHL2001 source numbers} \\ 
\noalign{\smallskip}\hline\noalign{\smallskip}
outside \xmm\ field: & 
1--24,27--33,35,37--41,46-48,50,52,54,58,60, &
1--32,34,35,37--43,45,47,50,52,53,55,57,58,60--63,\\
& 
61,65,70--72,78,79,81,85,87,88,91,95,98--100, &
66,67,69,74,75,77,78,80,81,83,85,87--91,93,98,100, \\
& 
102,104,110,114,120,122,123,125,129,134, &
103,106,107,109,113,115--120,123,125,127,128,130, \\
& 
135,137,148,154,159,169,171,180,183,185, &
131,133--135,141,145,146,149,155,158,160,164,165, \\
& 
199,202,209,212,216,219,224,227,230,236, &
167,176,179,181,185,186,189,192,196,201,202,204, \\
& 
237,242,245,246,257--260,264,265,268,269, &
205,210,215,220,221,224,225,228,233,234,239,243, \\
& 
271,273--276,282,285--287,290,293,294,298, &
245,248,251--253,259,260,263--265,270,271,274,276, \\
& 
300,305,312,314,318--320,327,335,339,341, &
278,280--282,284,290--293,295--298,300,302,305,309, \\
& 
342,350,356,358,362--365,367,371,373,374, &
311,313,317,325,326,328,336,355,358,361,364--367, \\
& 
376,377,381--383,385--396 &
369,371--373,375,377--379,381--383,387--396 \\
\noalign{\smallskip}
resolved by \xmm: & 
131,177,186,198,200,211,220,250,370 &
129,177,178,183,191,193,195,197,206,214,254,272,\\
& 
&
374 \\
\noalign{\smallskip}
not detected, LH$<$12: & (12 of 18 sources) &  (3 of 21 sources) \\
&
59,62,63,69,84,86,113,149,161,307,329,330   &
96,238,307\\
\noalign{\smallskip}
not detected, 12$\le$LH$<$15: & (9 of 29 sources) & (2 of 17 sources)\\
& 
49,82,93,109,128,196,283,334,372 &
76,79 \\
\noalign{\smallskip}
not detected, LH$\ge$15: &  (22 of 193 sources) & (18 of 167 sources)\\
& 
68(LH=14294),77(25),80(16),126(217), &
104(LH=901),121(94),171(43),173(317),190(216), \\
& 
133(40),160(26),166(17),190(113),191(55), &
207(98),208(299),230(76),232(1166),246(40),\\
& 
192(54),203(103),214(400),215(251), &
256(60),267(22),322(2703),324(148),344(41),\\
& 
232(104),263(38),270(40),277(16),284(20), &
356(15),380(17),384(16)\\
& 
309(82),325(21),331(20),340(28) &
\\
\noalign{\smallskip}
\hline
\noalign{\smallskip}
\end{tabular}
\label{rosat}
\end{center}
\end{table*}
In this section we discuss the cross-correlation of the \xmm\  detected sources
with sources reported in earlier X-ray catalogues. All correlations (together
with other X-ray information like variability (v) or transient nature (t), 
reported in these catalogues as well as
extent (ext) detected in this work) are indicated in the XID 
column of Table~2 (Col. 67).

From the 108 \ein\ sources reported in 
\citet[][hereafter TF91]{1991ApJ...382...82T}, 14 are outside  
the field covered by the \xmm\ observations
\germanTeX
([TF91]~6,""9,""11,""13,""15,""38,""81,""86,""95,""97,""98,""102,""106,""108). 
\originalTeX
From the remaining 94 sources
22 are not detected in our analysis while they were clear \ein\ detections 
\germanTeX
([TF91]~29,""30,""31,""35,""37,""39,""40,""43,""46,""47,""53,""65,""66,""72,""75,""78,""88,""93,""96,""99,""100,""107)
\originalTeX
and have to be classified as transient. Three \ein\ sources 
\germanTeX
([TF91]~10,""70,""103)
\originalTeX
are resolved by \xmm\ into two sources. 
Several of the transient candidates 
\germanTeX
([TF91]~31,""39,""40,""46,""47,""53,""75,""78,""96) were already classified as transients
\originalTeX
\germanTeX
or variable ([TF91] 37,""88)
\originalTeX
after the ROSAT HRI and PSPC observations 
\citep[][hereafter PFJ93, and SHL2001]{1993ApJ...410..615P}.
On the other hand, the \ein\ source [TF91] 84, classified as transient by
SHL2001, was detected with \xmm\ at an absorbed XID band luminosity of 1.5\ergs{36}, 
about a factor of 8 below the luminosity reported by TF91 and a factor of two below
the SHL2001 upper limit. This source therefore
either is a recurrent transient or just highly variable.

In the correlation with ROSAT detected sources we concentrate on the
catalogue of the central HRI pointing (PFJ93) performed in July 1990,
and the first (July 1991) and second (August 1992)
PSPC survey of the full galaxy (SHP97,SHL2001). With the HRI 86 sources
are detected within the central $\sim$34\arcmin\ of \m31. The \xmm\ observations 
cover the full field and detect all but 7 of the sources. The missing sources
\germanTeX
([PFJ93]~1,""2,""31,""33,""51,""63,""85) 
\originalTeX
are not in confused regions and covered the 
luminosity range (0.2--2.4)\ergs{37}. While 
\germanTeX
[PFJ93]~1,""2,""85 
\originalTeX
were still detected in
the first and/or second ROSAT PSPC survey, sources 
\germanTeX
[PFJ93]~31,""33,""51,""63 
\originalTeX
were no longer active.
All ROSAT HRI sources should have clearly been detected as bright sources 
in the \xmm\ survey and the missing sources therefore have to be classified as transients.   

The ROSAT PSPC surveys of \m31\ contain 396 sources each, covering 6.3 and
10.7 deg$^2$, an area much bigger than the 1.24 deg$^2$ of the \xmm\ survey.
In Table~\ref{rosat} we list sources of SHP97 and SHL2001 that are not detected by \xmm.
From ROSAT PSPC survey I and II, 156 and 191 sources are
outside the field covered by the \xmm\ observations, 43 and 23 are not detected 
by \xmm, respectively. For \me33\ we found in Paper I that many of the
sources with the lowest detection likelihoods LH were spurious detections. To
search for similar effects in the \m31 ROSAT catalogues we arranged the ROSAT
sources in the field, that are not detected by \xmm\ according to LH.
From the first ROSAT survey, we do not detect 12 out of 18 sources 
with LH below 12 and 9 out of
29 sources with LH below 15. These sources 
should have been detected by the deeper \xmm\
observations if still at the ROSAT brightness.  While in principle they all
could have dimmed by such an amount that we do not detect them, this seems
rather unlikely and probably most of these sources were spurious detections.
They may have originated if a too low background has been used for the long 
observations (exposures up to 50 ks) taken during the ``reduced pointing phase"
of the ROSAT mission. As Table~\ref{rosat} shows this problem is not affecting
the sources of the second ROSAT survey LH. 
There are many ROSAT PSPC sources detected with high LH that
are missing in our \xmm\ catalogue. Many of these sources have already been 
classified as transient comparing with the \m31\ \ein\ catalogue or just ROSAT 
survey I and II by SHP97 and SHL2001. At the position of one of these possible transients 
([SHL2001] 240) we detect a source with a luminosity of $\sim$1.5\ergs{35} 
(if located in \m31), which is about a factor of 100 fainter
than during the outburst detected by ROSAT.

There are \chandra\ catalogues of \m31\ X-ray sources based on observations of the centre area
with the ACIS-I \citep[204 sources, 43 not detected by \xmm,][]{2002ApJ...577..738K} and 
HRC \citep[142 sources, 18 not detected by \xmm,][]{2002ApJ...578..114K} detectors, 
and of shorter observations 
rastering part of the disk with the HRC 
\citep[166 sources, 4 outside \xmm\ field, 23 not detected by \xmm,][]{2004ApJ...609..735W}. 
While the  \chandra\  centre area surveys are fully covered by the \xmm\ survey, 
the \chandra\  disk raster partly exceeds
the field covered by the \xmm\ observations.
In addition to these larger \chandra\ survey papers, 
there are shorter lists of e.g. GlC sources and SSSs and reports of transients detected 
by \chandra. Correlations are given in the XID column of Table~2 and further 
references in the footnotes. While most of the \chandra\ detected sources can be 
resolved with \xmm\ EPIC, there are sources close to the \m31\ nucleus (319,321,325) 
or with arcsecond separation (360,366) that remain unresolved. Many of the \chandra\ sources
not detected by \xmm\ are rather faint (luminosities of a few \oergs{36} in the centre
area and down to \oergs{35} further out in the bulge) and therefore not resolved from the
surrounding diffuse emission. The remaining undetected \chandra\ sources are rather bright
and all but one of them (source n1-77 from \citet{2004ApJ...609..735W}) are already classified as transient 
\germanTeX
(s1-79,""s1-80,""r3-46,""s1-82,""r1-34,""r3-43,""r2-28,""r1-34,""s1-85,""r1-23,""r1-19,""n1-85).
\originalTeX

\chandra\ spatially resolved several SNRs in \m31. Only some of them are detected by
\xmm\ (see Sect. 6.4).

Only lists of bright \xmm\ sources were published up to now 
\citep[see e.g.][]{2001A&A...378..800O,TKP2004}. All of these sources
are contained in our catalogue. For bright sources in \m31,
many photons were collected by the \xmm\ EPIC cameras resulting in detailed  
light curves and spectra that could be analyzed in detail and allowed us to
define the source class (see references
in Table~2 and the discussion of source classes in the following section).  

\section{Classes of point-like X-ray sources detected in the direction of \m31}
\begin{figure}
   \resizebox{8.15cm}{!}{\includegraphics[bb=50 90 355 460,angle=-90,clip]{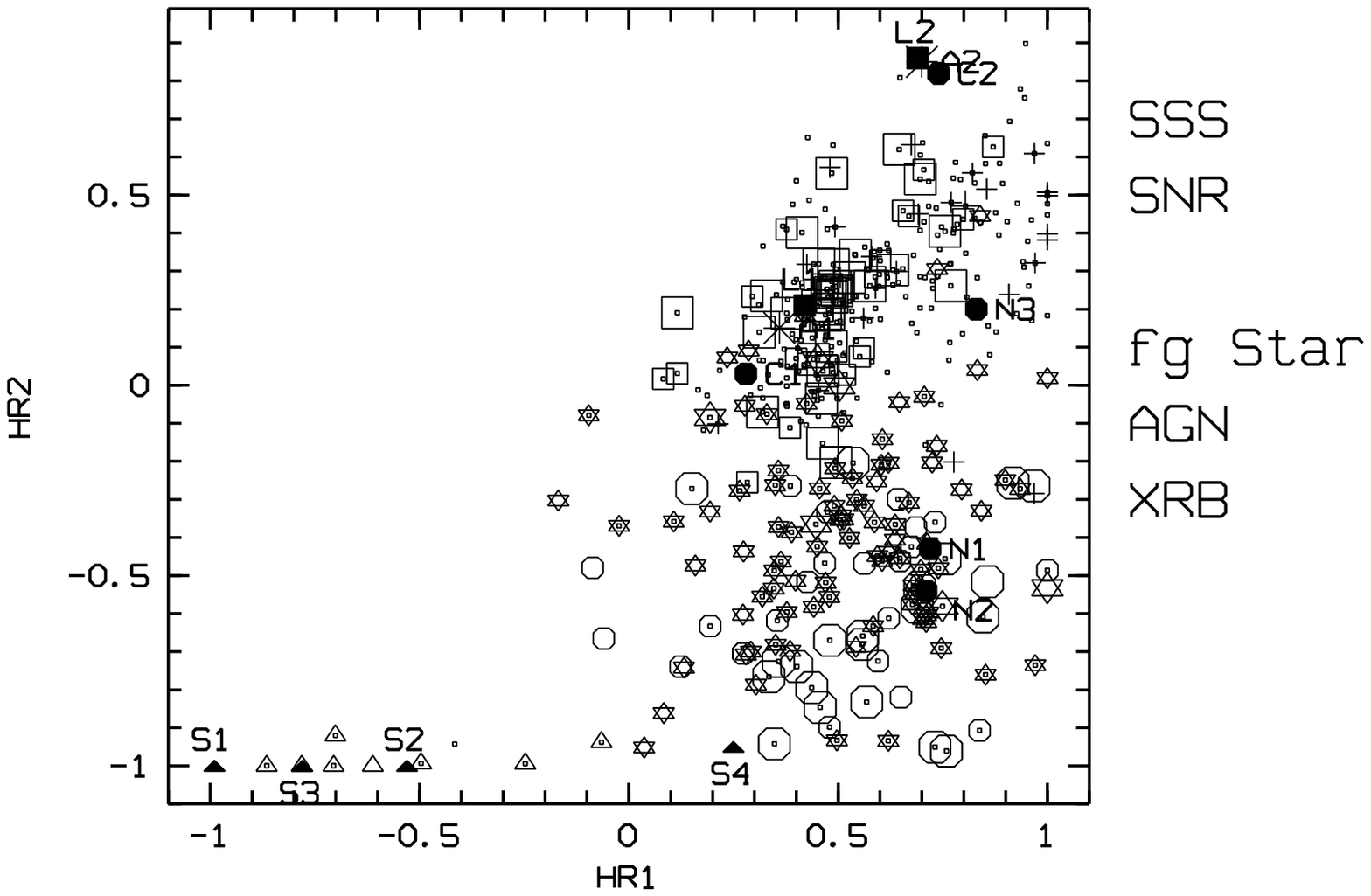}}
   \resizebox{8.15cm}{!}{\includegraphics[bb=50 90 340 460,angle=-90,clip]{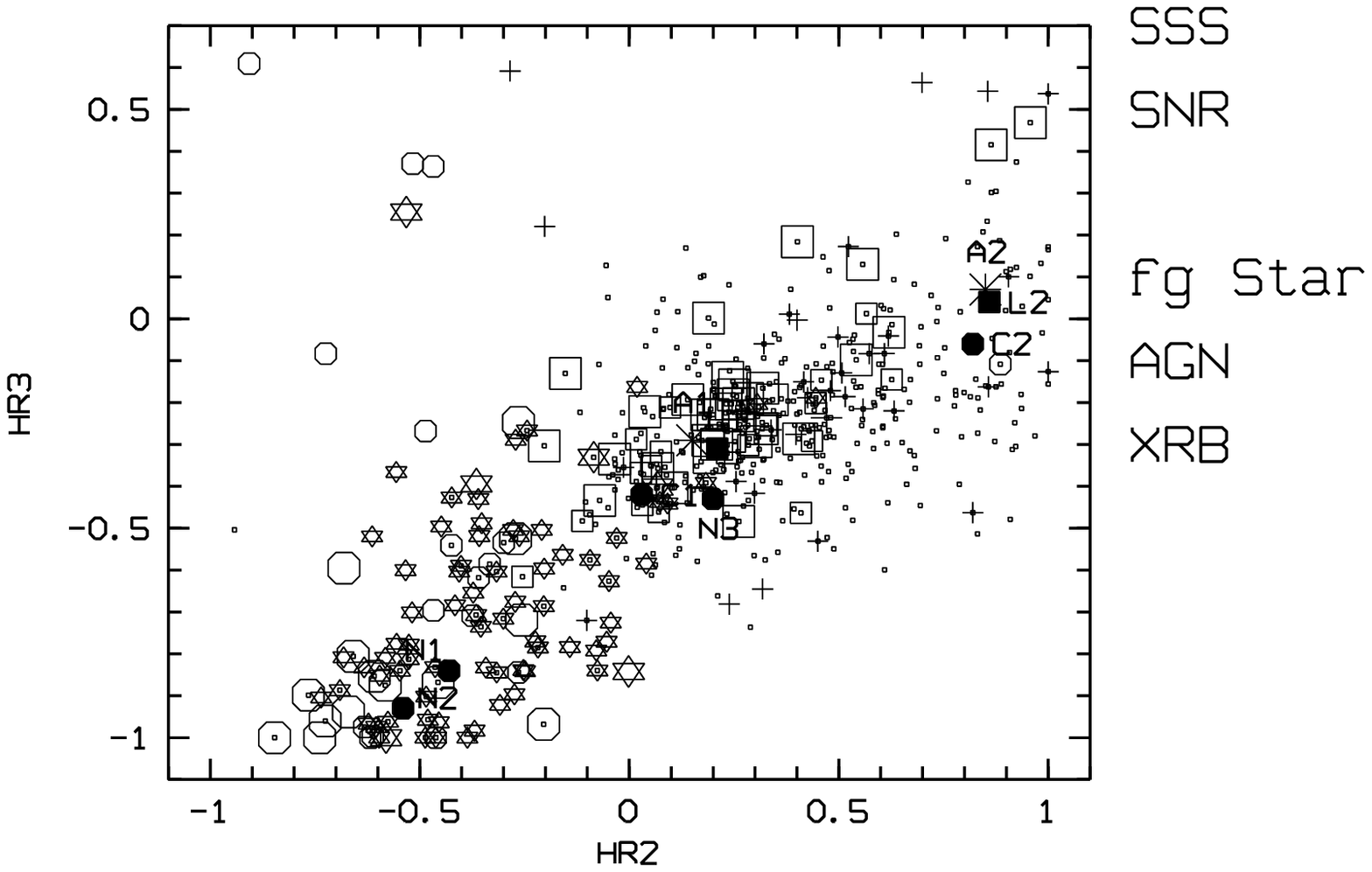}}
   \resizebox{8.15cm}{!}{\includegraphics[bb=50 90 355 460,angle=-90,clip]{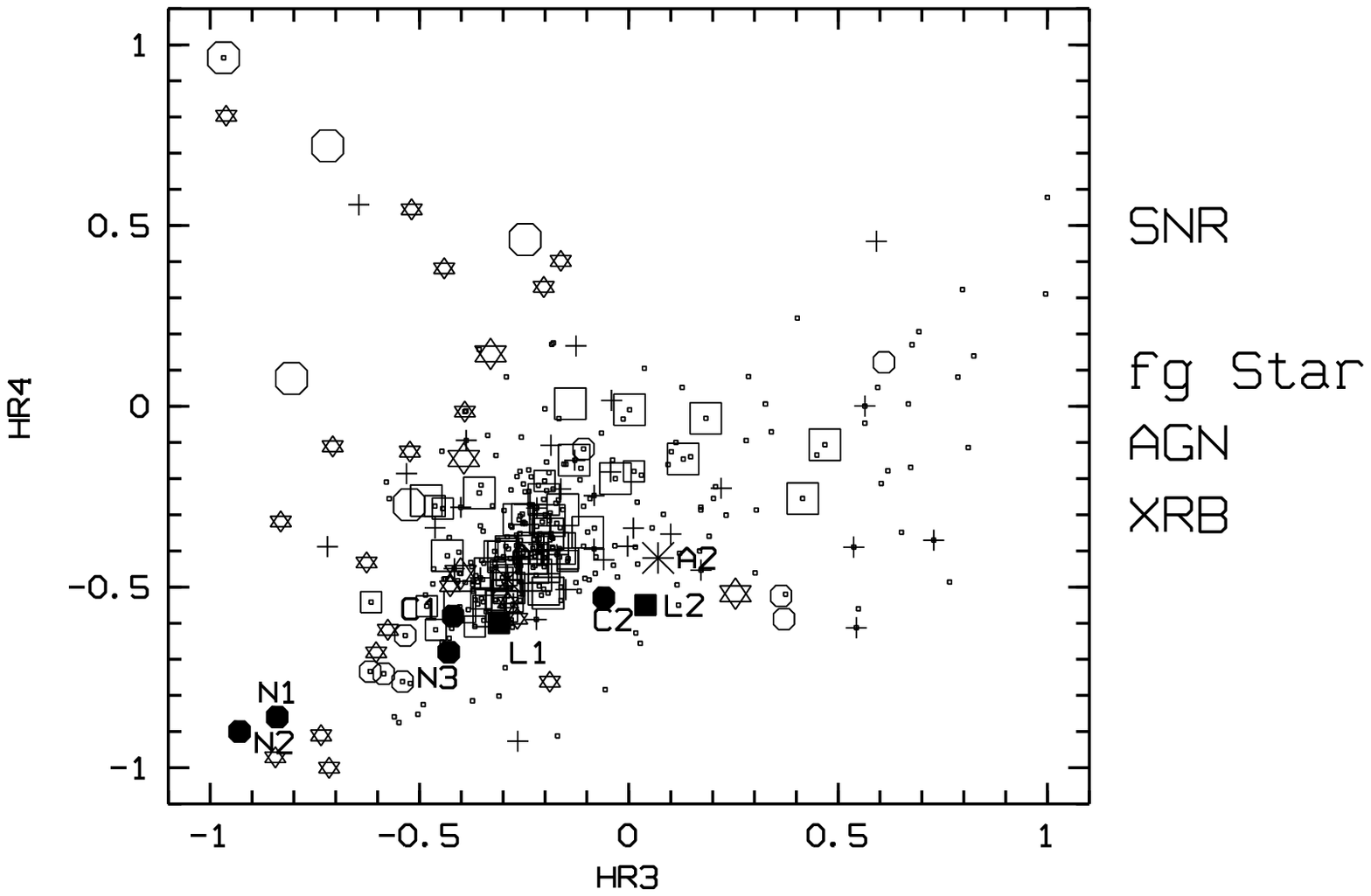}}
     \caption[]{
     Hardness ratios of sources detected by \xmm\ EPIC. Shown as dots are only
     sources with HR errors smaller than 0.20 on both $HR(i)$ and $HR(i+1)$. Foreground
     stars and candidate are marked as big and small stars, AGN and candidates as
     big and small crosses, SSS candidates as triangles, SNR and candidates as big and small
     hexagons, GlCs and XRBs and candidates as big and small squares. In addition, we mark positions derived from
     measured \xmm\ EPIC spectra and models  for SSSs (S1 to S4) as filled
     triangles, low mass XRBs (L1 and L2) as filled squares, SNRs (N132D as N1,
     1E~0102.2--7219 as N2, N157B as N3, Crab spectra as C1 and C2) as filled hexagons, 
     AGN (A1 and A2) as asterisk.
     }
    \label{hr} 
\end{figure}
\begin{table*}
\begin{center}
\caption[]{Summary of identifications and classifications.}
\begin{tabular}{llrr}
\hline\noalign{\smallskip}
\hline\noalign{\smallskip}
\multicolumn{1}{c}{Source type} & 
\multicolumn{1}{c}{Selection criteria} &
\multicolumn{1}{c}{identified} &
\multicolumn{1}{c}{classified}  \\ 
\noalign{\smallskip}\hline\noalign{\smallskip}
fg Star & ${\rm log}({{f}_{\rm x} \over {f}_{\rm opt}}) < -1.0$ and $HR2 - EHR2 < 0.3$ and $HR3 - EHR3 < -0.4$ or not defined & 6 & 90 \\
AGN  &  Radio source and not classification as SNR from $HR2$ or optical/radio & 1 & 36 \\
Gal  &  optical id with galaxy  & 1 &  \\
GCl  &  X-ray extent and/or spectrum &1 &1\\
SSS  &  $HR1 < 0.0$, $HR2 - EHR2 < -0.99$ or $HR2$ not defined, $HR3$, $HR4$ not defined &   &  18 \\
SNR  &  $HR1 > -0.1$ and $HR2 <-0.2$ and not a fg Star, or id with optical/radio SNR   & 21 & 23 \\
GlC  &  optical id & 27 & 10 \\
XRB  &  optical id or X-ray variability  & 7 & 9 \\
hard &  $HR2 - EHR2 > -0.2$ or only $HR3$ and/or $HR4$ defined, and no other classification&  & 567 \\
\noalign{\smallskip}
\hline
\noalign{\smallskip}
\end{tabular}
\label{class}
\end{center}
\end{table*}
To identify the X-ray sources in the \m31\ field we searched for correlations around the
X-ray source positions  within a radius of $3\times(\sigma_{stat} +
\sigma_{syst})$ in the SIMBAD and NED archives and within several catalogues. 
In columns 68 to 73 of Table~2, we give extraction information from
the USNO-B1  catalogue (name, number of objects within search area, distance,
B2, R2 and I magnitude of  the brightest object). 
To improve on the reliability
of identifications we have used the B and R magnitudes to calculate   ${\rm
log}({{\rm f}_{\rm x} \over {\rm f}_{\rm opt}}) = {\rm log}({f}_{\rm x}) +
({m}_{B2} + {m}_{R2})/(2\times2.5) + 5.37$,  following
\citet[][ see Col. 74]{1988ApJ...326..680M}.

The X-ray sources in the catalogue are identified  or classified based on properties
in the X-ray  (HRs, variability, extent) and of correlated
objects in other  wavelength regimes (Table~2, Cols. 75, 76).  The
criteria are summarized in Table~\ref{class} and are similar to the ones 
used in Paper I for \me33. However, one has to keep in mind, that \m31\ is more 
massive than \me33\ with more interstellar matter in the disk which in addition
is more inclined. Therefore sources in or behind 
\m31\ may suffer higher absorption than similar sources in the \me33\ field.
To take this into account, we compare our HRs to model HRs determined assuming 
Galactic foreground absorption and additional absorption within \m31\ of 
9\hcm{21}, a factor of 2 higher than in Paper I.  
We count a source as
{\em identified}, if at least two criteria secure the identification. Otherwise, we
only count a source as  {\em classified} (indicated by pointed brackets). 

We plot X-ray colour/colour diagrams based on the HRs (see Fig.~\ref{hr}).
Sources are plotted as dots if the error in both contributing HRs  is
below 0.2. Classified and identified sources are plotted as symbols even if 
the error in the contributing HRs is greater than 0.2. Symbols including a dot 
therefore mark the well defined HRs of a class.  
To identify areas of specific source classes in the plots, we
over-plot  colours of sources, derived from measured \xmm\ spectra and model
simulations.

Identification and classification criteria and results are discussed in detail
in  the subsections on individual source classes below. Many foreground stars,
SSSs and SNRs can be classified or identified. However, besides a few clearly
identified XRBs and AGN, and SNR candidates from positions in other
wavelengths, we have no clear hardness ratio criteria (see below) to select
XRBs, Crab-like SNR or AGN. They are all ``hard" sources and we therefore
introduced a class $<${\rm hard}$>$ for  sources with $HR2$ minus $HR2$ error greater
than $-$0.2. In the following subsections we first discuss foreground and background
sources in the catalogue and then sources within \m31.
Thirty-eight sources remain unidentified or without classification.

\subsection{Foreground stars (fg Star)}  
Foreground stars are a class of X-ray sources that is homogeneously distributed
over the FOV of \m31. The good positioning of \xmm\ and the 
available catalogues USNO-B1 and 2MASS allow us to effectively select this type of 
sources. All our foreground stars and candidates are 2MASS sources. 
We identified two sources (139 and 709) with stars of known type in the SIMBAD data
base. Two sources (552 and 565) are W UMa type variables identified in the DIRECT program 
\citep{1998AJ....115.1016K,1999AJ....117.2810S}. Two sources (284 and 385) were globular cluster 
candidates that have been identified as foreground stars \citep{2004A&A...416..917G}.
Their X-ray fluxes are in the range expected for these sources. In
addition, we classify 90 sources as foreground stars (${\rm log}({{f}_{\rm
x} \over {f}_{\rm opt}}) < -1$, see  \citet{1988ApJ...326..680M} and in
addition $HR2 - EHR2 < 0.3$ and $HR3 - EHR3 < -0.4$). 
For several of the star candidates we
estimate the type from the optical colours in the USNO-B1 catalogue using the stellar
spectral flux library from   \citet{1998PASP..110..863P}. For many candidates we
can not determine the type from the optical colours. This may indicate that these sources
are not isolated stars but more complicated systems or even in some cases
galaxies. This expectation can for instance be confirmed for source 766, which
is resolved into two sources in DSS2 images.

Several of the foreground star candidates close to the centre of \m31\   
(219,231,304,378,423,434) have no entry in the USNO-B1 catalogue,
however, they are clearly visible on Digital Sky Survey images, they are 2MASS
sources and fulfill the X-ray hardness ratio selection criteria. Therefore, we 
also classify them as foreground stars.  

For three sources (2,802,835) the HRs indicate a foreground star, however,  
${\rm log}({{f}_{\rm x} \over {f}_{\rm opt}})$ are (--0.98,--0.87,--0.81)  
slightly bigger than the assumed limit for classification. Comparing USNO-B1 and 
2MASS optical magnitudes for these sources clearly indicates variability. Therefore,
we still include them as foreground star candidates.

Searching for correlations in the SIMBAD database, we found that source 136 
correlates with the recurrent nova Rosino 61. \citet{1973A&AS....9..347R}
reports: ``maxima of 16.4 pg and 17.7 pg have been respectively observed 
on August 12, 1966 and Oct 25, 1968. In both cases, the star was near maximum 
for only one or two days, rapidly declining below the limit of visibility.
Although the apparent magnitude at maximum may be consistent with that of a 
nova at the distance of \m31, the light curve is abnormal, even for a 
recurrent nova. An alternative hypothesis is that the star may be not a 
nova, but a foreground U Gem variable, projected by chance over \m31."
The X-ray hardness ratios of the source and its 2MASS classification 
support the latter explanation and we therefore add the source as a 
foreground star candidate.  

For seven sources 
\germanTeX
(40,""78,""114,""206,""495,""562,""686), 
\originalTeX
${\rm log}({{f}_{\rm x} \over {f}_{\rm opt}})$ 
points at a stellar identification. 
However, the HRs including errors are outside the  assumed limits
for foreground stars. Source 686 in addition correlates with a radio source
making it a good candidate for an AGN behind \m31. 

Optical spectroscopy of all foreground star candidates is needed to prove
the suggested identification.

\subsection{Galaxies (Gal), galaxy clusters (GCl) and AGN}
Already after the \ein\ observations the X-ray bright source [TF91] 51 (315)  was
identified with the local group galaxy M~32. 
Mainly based on ASCA and ROSAT HRI observations, \citet{1998ApJ...497..681L}
argued that the X-ray emission is dominated by a LMXB, slightly offset from
the black hole nucleus of M~32. \chandra\ observations allowed to
resolve this central emission of M~32 into three sources, one coincident with
the nucleus and the by far brightest source at an 
offset from the galaxy nucleus of 8\farcs3 coincident with the position of 
the proposed LMXB \citep{2003ApJ...589..783H}.  
In addition to these sources, \citet{2004ApJ...609..735W} detected a bright 
X-ray transient in M~32 (s1-85). The \xmm\ source 315 is positioned far off-axis
in observation s1
and does not allow us to resolve the three \chandra\ sources close to the nucleus.
The luminosity of 5\ergs{37} and HRs of the source are certainly dominated by 
the LMXB and typical for this source class ($HR1=0.31, HR2=0.07, HR3=-0.28,
HR4=-0.33$). 
The M~32 
transient detected by \chandra, was not active during the \xmm\ observation.

The brighter of the two extended sources in our survey (747) has been identified 
as a cluster of galaxies (GCl) at a redshift of z=0.293 based on the X-ray spectrum 
\citep{KTV2003}. The second extended source (832) is detected in the northeast
part of the n3 field outside the optical $D_{25}$ ellipse of \m31. We therefore 
classify this source as a GCl candidate.

There are no correlations with AGN with known redshift. However, the brightest
X-ray source in the \m31\ field (50) which was always active since the first \ein\
observations correlates with a radio source and an unresolved optical object
with a magnitude in the R band of 18.2. It has been identified as a BL Lac type AGN
(NED).   

In addition, we classify 36 sources as AGN based on SIMBAD, NED, and other radio 
source correlations (NVSS, \citet{1990ApJS...72..761B},
\citet{2004ApJS..155...89G}) 
with the additional condition of being not a SNR or SNR candidate from the X-ray 
HR. A final decision will only be possible from 
optical spectra as some of the sources may still be plerion type SNRs. 
In Fig.~\ref{hr} we include typical HRs for an AGN spectrum
(power law with photon index of 1.7 assuming Galactic foreground absorption
(A1) and absorption of 9\hcm{21} through \m31\ (A2)). The AGN candidates
populate the area in the HR diagrams expected from the model spectra. Many of
the other sources in that HR diagram area -- now just classified as hard --
may turn out to be AGN.

\subsection{Super-Soft Sources (SSSs)}
SSSs show black body spectra with temperatures below 50 eV, radiate close to the
Eddington luminosity of a 1 $M_{\sun}$ object and are believed to be white dwarf
systems steadily burning hydrogen at the surface. They were identified as a
class of X-ray sources by ROSAT and are often observed as transient X-ray
sources  \citep[see][ and references therein]{2000NewA....5..137G}. In the
catalogue, SSSs are only classified using their HRs.  To guide the classification
we calculated SSS HRs assuming a 25  and 50 eV black body spectrum assuming
Galactic foreground  absorption (S1 and S3) and absorption of 9\hcm{21} within 
\m31\  (S2 and S4). At the high absorption, the sources would no longer be 
detectable with \xmm\ as the intrinsic flux (expected to be around  \oergs{38})
would be reduced by 5 and 4 orders of magnitude for the two model spectra, 
respectively.   
This led to the selection criteria in
Table~\ref{class} and the classification of 18 \m31\ SSSs. 
They are detected  with absorbed XID band luminosities of
6\ergs{34} to 9\ergs{36}. Only eight of the sources and the
model values can be plotted in Fig.~\ref{hr}.  For the other 10, $HR2$ is not
determined. 

One of the SSS candidates (430) was already detected by \ein\ and
ROSAT. It was found to be variable, however not classified as a SSS 
in the ROSAT survey which may
be caused by nearby sources unresolved by the ROSAT PSPC. This source
as well as another six 
\germanTeX
(313,""320,""336,""359,""369,""431) 
\originalTeX
were also detected by \chandra. Sources 
\germanTeX
320,""336,""367,""431 
\originalTeX
were classified as supersoft transients 
\citep{2004ApJ...609..735W,2004ApJ...610..247D}.
Source 431 shows a 865s period in the \xmm\ data reminiscent of 
a rotation period of a magnetized white dwarf \citep{2001A&A...378..800O}.
Four of the sources correlate with optical novae 
\germanTeX
(313,""347,""359,""456)
\originalTeX
and two correlate with 2MASS sources, indicating a dusty surrounding (191,401). 

A special case is source 352, where $HR1$ indicates a SSS, however,
$HR2$ is 0.94 and within the error not compatible with --1.0 and $HR3$ and $HR4$ are not 
undefined. Therefore we do not classify the source as a SSS. On the other hand, 
this source was classified as a variable SSS from \chandra\ observations 
\citep{2004ApJ...610..247D}. It is detected by \ein\ and ROSAT.
There are several possibilities: Due to the position close to the centre of \m31\
and the high source density in this region, \xmm\ may not be able to resolve this 
source from nearby sources, the source detection may be fooled by the surrounding 
diffuse emission, or it is not a source with typical SSS properties.

Another source (395), classified as a transient SSS from \chandra\ observations 
\citep[][]{2004ApJ...609..735W,2004ApJ...610..247D}, has $<$hard$>$ HRs in
\xmm\ and is classified here as XRB candidate.

In the discussion above already two sources (352,395) were mentioned that are 
also covered by the \chandra\ survey for SSSs and QSSs. 
In this survey, \citet{2004ApJ...610..247D} determined 33 SSSs and 
QSSs using rates and hardness ratio criteria based on the \chandra\ ACIS-S 
(0.1--1.1), (1.1--2), and (2--7) keV bands. As the authors discuss, this survey
not only selects classical SSSs as defined above but also includes sources with effective
temperatures as high as 350 eV. This expectation is confirmed by the 16 sources 
of the catalogue which coincide with
sources from the \xmm\ catalogue. Here, three of them are classified as SSS candidates
(320,336,369), one as SNR (154), three as SNR candidates (295,318,704), four as
foreground star candidates (128,157,231,727), two as XRB candidates (321,395),
and one even as hard (169).

\subsection{Supernova remnants (SNR)}
SNRs can be separated in sources where thermal components dominate the X-ray
spectrum below 2 keV (examples are N132D in the Large Magellanic Cloud and
1E~0102.2--7219 in the Small Magellanic Cloud) and  so called ``plerions" with
power law spectra (examples are the Crab nebula and N157B in the Large
Magellanic Cloud). To guide the classification we calculated HRs from archival
\xmm\ spectra of these SNRs. Spectra of N132D (N1),  1E~0102.2--7219 (N2), and
N157B (N3)  can directly be compared to \m31\ SNRs as they are seen through 
comparable foreground absorption. For the Crab nebula spectrum, we assumed 
Galactic foreground absorption (C1) and absorption of 9\hcm{21} within 
of \m31\ (C2). It is clear from Fig.~\ref{hr}, that ``thermal" SNRs are located
in areas of the X-ray colour/colour diagrams that only overlap with foreground stars. If we
assume that we have identified all foreground star candidates from the optical
correlation and inspection of the optical images, the remaining sources can be
classified as SNRs with the criteria given in  Table~\ref{class}. Compared to 
our \me33\ analysis in Paper I, we relaxed our $HR2$ selection a bit to adjust to 
the higher absorption in the \m31\ disk.

Extensive searches for SNRs in \m31\ were performed in the optical 
\citep[ and references therein]{1980A&AS...40...67D,1995A&AS..114..215M} and radio
\citep{1993A&AS...98..327B}. Before \chandra\ and \xmm, \m31\ X-ray sources were
identified as SNR just by positional coincidence with optical and radio catalogues. 
Also for our catalogue, many X-ray sources
correlate with optical and/or radio SNRs or radio sources (see
Table~2). With \chandra\ and \xmm, we now have the possibility to identify
\m31\ SNRs by their X-ray properties (extent, HRs, or spectra).
\chandra\ observations of sources 260, 342 and 454 
\citep[][]{2002ApJ...580L.125K,2004ApJ...609..735W} were analyzed in detail
and the SNR nature determined by X-ray extent and optical/radio correlation. 
We count sources as SNR identifications (21) if they correlate with optical 
and/or radio SNRs or radio sources  
and fulfill the HR criteria (see above). We count sources as classified
SNRs (23) if they either  fulfill the HR criteria (22) or correlate with an 
optical/radio SNR (1). 
The SNRs and candidates cover an absorbed XID band luminosity range 
from 4\ergs{34} to 5\ergs{36}. 
The 22 HR selected SNR candidates have just been selected from their X-ray
properties and significantly enlarge the number of \m31\ SNR candidates. For
three of these sources, the classification may be questionable: 
From \chandra\ observations \citep{2002ApJ...580L.125K}, 
source 316 is classified as variable, and
295 and 318 correlate with a star of the catalogue by \citet{1994A&A...286..725H}
.

A special case is source 642 with an XID band luminosity of 1.0\ergs{36}, 
which correlates with the optical SNR candidate 
[MPV95] 1-013  \citep{1995A&AS..114..215M}. The \xmm\ HRs clearly point at a 
hard spectrum as expected for a plerion embedded in the \m31\ disk.
If the identification is correct this source would be the first plerion detected 
outside the Galaxy or the Magellanic Clouds. Final proof can only be achieved if
the source turns out to be extended and shows no time variability.

\citet{2003ApJ...590L..21K} reported the \chandra\ detection of two additional 
resolved X-ray SNRs  in 
the centre of \m31. The first of the sources lies about 40\arcsec\ northeast of the 
\xmm\ source pair 338,341, the second inbetween these two sources. With X-ray luminosities of
(4 and 8)\ergs{35} in the (0.3--7) keV band, they are too faint to be detected by \xmm\ 
in the neighbourhood of the bright sources and surrounding diffuse emission.  

\subsection{Globular cluster sources (GlC)}
A significant part of the luminous X-ray sources in the Galaxy and \m31\ are found 
in globular clusters. Most of the GlCs in the Galaxy show bursts and therefore are
low mass neutron star X-ray binaries \citep[see e.g.][]{VL2004}. There have 
been extensive surveys for globular clusters in \m31\ in the optical and infrared
band \citep[see Magnier 1993,][]{2004A&A...416..917G}. From the ROSAT survey sources,
SHL2001 identify 33 sources as GlCs. We correlated the \xmm\ catalogue with the catalogues of
\citet{2004A&A...416..917G} and Magnier (1993) and add one candidate, found in SIMBAD. 
We count sources as identifications (27) when the correlating source is listed as 
confirmed GlC in \citet{2004A&A...416..917G}. The remaining correlations are
counted as classified. Two sources (91,146), correlating with GlC candidates in  
\citet{2004A&A...416..917G} showed X-ray HRs typical of SNRs, 
source 91 correlates with 
an optical SNR candidate and both with radio sources. We therefore identify
them as SNRs. 

In Fig.~\ref{hr}, the GlCs are plotted with the same symbols as the XRBs.
All of our GlC identifications and candidates have been reported in earlier work 
and many of them have been classified as time variable in X-rays (see  
Table~2). They cover an absorbed XID band luminosity range 
from 4.5\ergs{35} to 2.4\ergs{38}. Only the brightest
source (351) has a luminosity that seems to be at the upper end, allowed for neutron 
star LMXBs. 

For source 414, \citet{2002ApJ...581L..27T} report intensity dips with a 
period of 2.78 hr reminiscent of a neutron star LMXB.

\subsection{X-ray binaries (XRB)}
As already mentioned in the introduction to this section,  expected spectra of
XRBs are similar to AGN and Crab-like SNRs. To guide the classification we
calculated HRs as expected for low mass XRBs (5 keV thermal Bremsstrahlung
spectrum  assuming Galactic foreground  absorption (L1) and absorption of 
9\hcm{21} within \m31\ (L2)). High mass XRB (HMXB) spectra are expected to 
be even harder. As can be
seen in Fig.~\ref{hr} the different source classes do not separate. 

Detailed work on individual sources 
using \xmm\ and/or \chandra\ data, has identified
four black hole XRBs \citep[sources 257,287,310,384,][]{2003A&A...405..505B,
2001ApJ...563L.119T,2004A&A...423..147B},
three neutron star LMXBs \citep[297,353,403,][]{2003A&A...405..505B,
2002ApJ...578..114K,2004A&A...419.1045M} and an XRB pulsar \citep[544][]{TKP2004}. 
In addition, we classify X-ray transients from \xmm\ and/or 
\chandra\ data as XRB candidates. While in general transient behaviour of
bright X-ray sources indicates an XRB nature, in this selection we may also pick 
up variable background sources. Such a source may be (405) which also correlates
with a radio source. The XRBs selected in this way cover
the same area  in the X-ray colour/colour diagrams as the GlCs (see Fig.~\ref{hr}). The absorbed XID band luminosities range from 
8.4\ergs{35} to 2.8\ergs{38}.

Besides the \chandra\ transients  covered in the \xmm\ catalogue, 
we identified many transients in earlier
catalogues in Sect. 5 that are further XRB candidates.
There certainly are more transients XRBs buried in the \xmm\ catalogue which only
could be discovered by a detailed comparison of fluxes of all the \xmm\ sources with 
earlier missions. Such an analysis was
outside the scope of this work and, due to the lower sensitivity of
the catalogues before \xmm\ and \chandra, transient sources detected on this 
basis may be less reliable. On the other hand, also X-ray sources classified as 
variable may turn out to be XRBs.
However, this criterion is not unique as it would also cover many variable background sources.
Up to now no high mass XRBs are identified in \m31. Several identifications with emission line 
objects (EmO) in SIMBAD may be good candidates for Be-type high mass XRB.

All the candidates mentioned above need further X-ray work to confirm an  XRB
nature.

\section{Summary}
We present a catalogue of 856 X-ray sources based on archival \xmm\ observations covering 
an area of 1.24 square degree.
We correlate our sources with earlier \m31\ X-ray catalogues and 
use information from optical, infra-red and 
radio wavelengths and in addition their X-ray properties for source classification.
We applied a source classification scheme similar to the one successfully used in Paper I 
for \me33.   
As \m31\ sources we detect 21 SNRs and 23 SNR candidates,
18 SSS candidates, 7 XRBs and 9 XRB candidates, as well as 27  
GlCs and 10 GlC candidates, which most likely are low mass 
XRBs within the GlC. Comparison to earlier X-ray surveys reveals transients not 
detected with \xmm, which add to the number of \m31\ XRBs.
The number of 44 SNRs and 
candidates more than 
doubles the X-ray detected SNRs. 22 sources are new SNR candidates in \m31\  
based on X-ray selection criteria.
Another SNR candidate may be the first plerion detected outside the Galaxy and the
Magellanic Clouds.
On the other hand,
six sources are foreground stars and 90 foreground star candidates, 
one is a BL Lac type AGN and 36 are AGN candidates, 
one source coincides with the Local Group galaxy M 32, 
one with a background galaxy cluster and another is a GCl 
candidate, all sources not connected to \m31.

There are 567 sources classified as hard, which may either be 
XRBs or Crab-like SNRs in \m31\ or background AGN. If source variability 
is detected SNRs may be excluded from source identification.
Thirty-eight sources remain unidentified or without classification.
  
The archival \xmm\ \m31\ observations allowed us to probe the
point source population of the covered area significantly deeper than during
the ROSAT PSPC surveys. However, the \xmm\ EPIC coverage is rather inhomogeneous and 
a significant part of the \m31\ disks has not been observed at all. To get a full 
census of the \m31\ X-ray point sources it would be very important to fully
cover this nearby galaxy with \xmm\ observations.

\begin{acknowledgements}
This publication makes use of the USNOFS Image and Catalogue Archive
operated by the United States Naval Observatory, Flagstaff Station
(http://www.nofs.navy.mil/data/fchpix/), 
of data products from the Two Micron All Sky Survey, 
which is a joint project of the University of Massachusetts and the Infrared 
Processing and Analysis Center/California Institute of Technology, funded by 
the National Aeronautics and Space Administration and the National Science 
Foundation, of the SIMBAD database,
operated at CDS, Strasbourg, France, 
and of the NASA/IPAC Extragalactic Database (NED) 
which is operated by the Jet Propulsion Laboratory, California 
Institute of Technology, under contract
with the National Aeronautics and Space Administration.
The \xmm\ project is supported by the Bundesministerium f\"{u}r
Bildung und Forschung / Deutsches Zentrum f\"{u}r Luft- und Raumfahrt 
(BMBF/DLR), the Max-Planck Society and the Heidenhain-Stiftung.
\end{acknowledgements}

\bibliographystyle{aa}
\bibliography{./paper_acc,/home/wnp/data1/papers/my1990,/home/wnp/data1/papers/my2000,/home/wnp/data1/papers/my2001}

\end{document}